\documentclass{pasa}%

\jid{PASA}
\doi{10.1017/pas.\the\year.xxx}
\jyear{\the\year}

\usepackage{graphicx,rotating}
\usepackage{siunitx} 

\usepackage{aas_macros}
\usepackage{hyperref} 
\hypersetup{colorlinks,citecolor=blue,linkcolor=blue,urlcolor=blue}
\usepackage{gensymb}
\usepackage{enumitem}

\def\swift{{\em Swift}}
\def\fermi{{\em Fermi}}
\def\robbie{{\sc Robbie}}
\def\aegean{{\sc Aegean}}
\def\grb{GRB\,180805A}
\def\JyPerBeam{{\rm Jy\,beam^{-1}}}

\def\ASU{$^{1}$}
\def\Curtin{$^{2}$}
\def\CIRA{$^{3}$}
\def\USydney{$^{4}$}
\def\UToronto{$^{5}$}

\def\UWisc{$^{7}$}
\def\UW{$^{8}$}
\def\UWA{$^{9}$}
\def\INAF{$^{10}$}
\def\CAASTRO{$^{11}$}

\author[]{
A.~P.~Beardsley\ASU,
B.~Crosse\Curtin,
D.~Emrich\CIRA,
T.~M.~O.~Franzen\Curtin,
B.~M.~Gaensler\USydney$^,$\CAASTRO$^,$\UToronto,
L.~Horsley\CIRA,
M.~Johnston-Hollitt\Curtin,
D.~L.~Kaplan\UWisc, 
D.~Kenney\Curtin,
M.~F.~Morales\UW, 
D.~Pallot\UWA,
K.~Steele\CIRA,
S.~J.~Tingay\Curtin$^,$\CAASTRO$^,$\INAF, 
C.~M.~Trott\Curtin$^,$\CAASTRO,
M.~Walker\CIRA,
R.~B.~Wayth\Curtin$^,$\CAASTRO, 
A.~Williams\CIRA, 
C.~Wu\UWA
\\
$^{1}$School of Earth and Space Exploration, Arizona State University, Tempe, AZ 85287, USA\\
$^{2}$International Centre for Radio Astronomy Research, Curtin University, Bentley, WA 6102, Australia\\
$^{3}$Curtin Institute of Radio Astronomy, Curtin University, GPO Box U1987, Perth WA 6845\\
$^{4}$Sydney Institute for Astronomy, School of Physics, The University of Sydney, NSW 2006, Australia\\
$^{5}$Dunlap Institute for Astronomy and Astrophysics, University of Toronto, ON, M5S 3H4, Canada\\
$^{6}$School of Chemical \& Physical Sciences, Victoria University of
Wellington, Wellington 6140, New Zealand\\
$^{7}$Department of Physics, University of Wisconsin--Milwaukee,
Milwaukee, WI 53201, USA\\
$^{8}$Department of Physics, University of Washington, Seattle, WA 98195, USA\\
$^{9}$International Centre for Radio Astronomy Research, University of Western Australia, Crawley 6009, Australia\\
$^{10}$Istituto Nazionale di Astrofisica (INAF) -- Istituto di Radio
Astronomia, Via Piero Gobetti, Bologna, 40129, Italy\\
$^{11}$ARC Centre of Excellence for All-sky Astrophysics (CAASTRO)\\
}

\title[MWA triggering]{Murchison Widefield Array rapid-response observations of the short GRB 180805A}

\author[Anderson et al.]{G.~E.~Anderson$^{1}$\thanks{E-mail: gemma.anderson@curtin.edu.au}, 
P.~J.~Hancock$^{1}$,
A.~Rowlinson$^{2,3}$, 
M.~Sokolowski$^{1}$, 
A.~Williams$^{1}$, 
J.~Tian$^{1}$, 
J.~C.~A.~Miller-Jones$^{1}$, 
N.~Hurley-Walker$^{1}$, 
K.~W.~Bannister$^{4}$, 
M.~E.~Bell$^{5}$, 
C.~W.~James$^{1}$, 
D.~L.~Kaplan$^{6}$, 
Tara~Murphy$^{7}$, 
S.~J.~Tingay$^{1}$, 
B.~W.~Meyers$^{1,8}$,
M.~Johnston-Hollitt$^{1}$,
R.~B.~Wayth$^{1}$
\affil{$^1$International Centre for Radio Astronomy Research, Curtin University, GPO Box U1987, Perth, WA 6845, Australia}
\affil{$^2$Anton Pannekoek Institute, University of Amsterdam, Postbus 94249, 1090 GE, Amsterdam, The Netherlands}
\affil{$^3$Netherlands Institute for Radio Astronomy (ASTRON), PO Box 2, 7990 AA Dwingeloo, The Netherlands}
\affil{$^4$Australia Telescope National Facility, CSIRO Astronomy and Space Science, PO Box 76, Epping, NSW 1710, Australia}
\affil{$^5$University of Technology Sydney, 15 Broadway, Ultimo NSW 2007, Australia}
\affil{$^6$Department of Physics, University of Wisconsin-Milwaukee, 1900 E. Kenwood Boulevard, Milwaukee, WI 53211, USA}
\affil{$^7$Sydney Institute for Astronomy, School of Physics, The University of Sydney, NSW 2006, Australia}
\affil{$^{8}$Department of Physics \& Astronomy, University of British Columbia, 6224 Agricultural Road, Vancouver, BC V6T 1Z1, Canada}
}%

\begin{document}

\begin{frontmatter}
\maketitle

\begin{abstract}
Here we present stringent low-frequency (185\,MHz) limits on coherent radio emission associated with a short-duration gamma-ray burst (SGRB).
Our observations of the short gamma-ray burst (GRB) 180805A were taken with the upgraded Murchison Widefield Array (MWA) rapid-response system, which triggered within 20\,s of receiving the transient alert from the \swift{} Burst Alert Telescope, corresponding to 83.7\,s post-burst.
The SGRB was observed for a total of 30\,min, resulting in a $3\sigma$ persistent flux density upper-limit of 40.2\,mJy\,beam$^{-1}$. 
Transient searches were conducted at the \swift{} position of this GRB on 0.5\,s, 5\,s, 30\,s and 2\,min timescales, resulting in $3\sigma$ limits of $570-1830$, $270-630$, $200-420$, and $100-200$ mJy\,beam$^{-1}$, respectively.
We also performed a dedispersion search for prompt signals at the position of the SGRB with a temporal and spectral resolution of 0.5\,s and 1.28\,MHz, respectively, resulting in a $6\sigma$ fluence upper-limit range from 570\,Jy\,ms at DM$=3000$\,pc\,cm$^{-3}$ ($z\sim 2.5$) to 1750\,Jy\,ms at DM$=200$\,pc\,cm$^{-3}$ ($z\sim 0.1)$, corresponding to the known redshift range of SGRBs.
We compare the fluence prompt emission limit and the persistent upper-limit to SGRB coherent emission models assuming the merger resulted in a stable magnetar remnant. 
Our observations were not sensitive enough to detect prompt emission associated with the alignment of magnetic fields of a binary neutron star just prior to the merger, 
from the interaction between the relativistic jet and the interstellar medium (ISM) 
or persistent pulsar-like emission from the spin-down of the magnetar.  
However, in the case of a more powerful SGRB (a gamma-ray fluence an order of magnitude higher than GRB 180805A and/or a brighter X-ray counterpart), our MWA observations may be sensitive enough to detect coherent radio emission from the jet-ISM interaction and/or the magnetar remnant.
Finally, we demonstrate that of all current low frequency radio telescopes, only the MWA has the sensitivity and response times capable of probing prompt emission models associated with the initial SGRB merger event.

\end{abstract}

\end{frontmatter}

\section{Introduction}
\label{sec:intro}

The detection of gravitational waves from the binary neutron star (BNS) merger GW 170817, and its association with the short-duration gamma-ray burst (GRB) 170817A, heralds a new era of observational astrophysics \citep{abbott17a,abbott17b}.
This discovery demonstrated the importance of rapid multi-wavelength follow-up, not only for providing complementary insight into the central engine, the energy released, and the final merger remnant, but primarily for localising the gravitational wave (GW) event. 
However, the unconstrained localisation capability of the dedicated gravitational wave detectors such as Advanced LIGO and Virgo \citep[aLIGO/Virgo; up to thousands of square degrees, eg][]{Abbott2016} still limits the efficiency of even the widest field-of-view instruments to identify the electromagnetic counterpart. 

In order to improve our understanding of the early-time (seconds to minutes) electromagnetic counterparts to gravitational wave events involving BNS and neutron star and black hole (NS-BH) mergers, we have devised a program to exploit their association with short-duration GRBs (SGRBs).
This is achieved by using the upgraded Murchison Widefield Array \citep[MWA,][]{tingay13,wayth18} rapid-response system \citep{hancock19b}, which allows the MWA to respond to SGRB transient alerts, such as those broadcast by the \textit{Neil Gehrels Swift Observatory} \citep[hereafter \swift;][]{gehrels04} Burst Alert Telescope (BAT) and the \textit{Fermi Gamma-ray Space Telescope} (hereafter \fermi{}) Gamma-ray Burst Monitor \citep[GBM;][]{meegan09}.
Using this system, the MWA can automatically observe the event within seconds of its discovery \citep[e.g. GRB 150524A;][]{kaplan15}.
As SGRBs are usually better localised than GW events, performing triggered observations on SGRBs provides the fastest and most efficient method for localising GW radio signatures, bypassing the need for tiling large areas of sky \citep[e.g.][]{kaplan16} or targeting a sample of nearby galaxies \citep[e.g.][]{hallinan_radio_2017}.
Such triggered observations will provide a template for the expected radio brightness and timing properties of GW mergers, which will in-turn inform the follow-up of future aLIGO/Virgo GW events by wide-field instruments like the MWA, the Australian Square Kilometre Array Pathfinder \citep[ASKAP;][]{hotan_australian_2014} and the low frequency component of the Square Kilometre Array (SKA-Low).%

On the very earliest timescales (seconds to minutes post-burst), merging BNS and NS-BHs are expected to produce prompt, fast radio burst (FRB)-like signals.
The existence of repeating FRBs \citep{Spitler2016,CHIME2019b,CHIME2019c,CHIME2020a} and a high volumetric rate \citep[$\gtrsim 10^4$\,Gpc$^{-3}$\,yr$^{-1}$; ][]{Ravi2019} argue against SGRBs being the sole progenitors of FRB. 
However, it is entirely possible that observed FRBs are produced by two populations of progenitors \citep{Caleb2019repeaters}, or that repeat bursts are produced by the orbital interactions of compact objects which later merge in an SGRB \citep[e.g.][]{Wang2016Inspirals}.
Indeed, the volumetric rate of bright FRBs --- only two of which have been observed to repeat despite significant follow-up observations \citep{Kumar2019,Kumar2020,James2020a} --- is remarkably consistent with that of the SGRB population \citep{james19c}.

Several models exist that predict coherent radio emission either just prior, during, or shortly following a SGRB/BNS/NS-BH merger. 
For example, prompt FRB-like emission is predicted to be produced just prior to the merger by magnetic braking as the magnetic fields of the NSs are synchronised to the binary rotation \citep{hansen01,lyutikov13}. 
A low-frequency radio burst could also be generated when the potential GRB jet first interacts with the surrounding interstellar medium \citep[ISM; see][]{usov00}. 
Following the merger, a supramassive, rapidly rotating and highly magnetised NS, often referred to as a magnetar, may be produced before either becoming stable or further collapsing into a black hole within seconds to hours  \citep{usov92,rowlinson13}. 
The remnant magnetar may produce pulsar-like radio pulses for its lifetime, which steadily fade as it spins down through the release of dipole radiation.
As it is expected that such emission will be beamed along the magnetic axis (as is the case for standard pulsars), it can be argued that this emission will be aligned with the relativistic jet axis that produced the SGRB, meaning it may remain preferentially pointed towards Earth and therefore observable as persistent radio emission \citep{totani13}. 
In the event of an unstable magnetar, an FRB could be produced through the ejection of its magnetosphere as it collapses into a black hole \citep{falcke14,zhang14}. This collapse could occur between 2 minutes to 3 hours post-merger depending on the mass, spin, and spin-down-rate of the remnant and the nuclear equation-of-state \citep{ravi14}. 
The detection of such prompt radio emission from SGRBs would allow us to distinguish between different binary merger models and scenarios, which would in-turn constrain the equation-of-state of nuclear matter \citep{lasky14}. For a review of the coherent radio emission mechanisms from compact binary mergers see \citet{rowlinson19}.

Radio signals are delayed as they propagate through astrophysical plasmas, with low-frequency signals arriving later than those at high frequencies.
This delay gives low-frequency telescopes such as the MWA the potential to catch a radio signal emitted simultaneously, or even before, the SGRB \citep{inoue04}, provided they are capable of rapidly responding to transient alerts.
The rapid-response (triggering) mode on the MWA enables us to study the very earliest low-frequency signatures from GRBs as it can be on target within seconds of an alert \citep[e.g. the MWA was on-target 23s following the \swift{}-detected SGRB 150424A;][]{kaplan15}. Since the second half of 2018, the MWA has been running an updated rapid-response mode that is capable of performing triggered observations on VOEvents, which are the standard for broadcasting alerts related to transient events \citep{seaman11}. 
Using the `4 Pi Sky VOEvent Broker' software \citep{staley16pp} and the Comet VOEvent client \citep{swinbank14}, we are able to monitor, parse and filter the VOEvent streams from both \swift{} and \fermi{}, allowing for triggers to be automatically pushed to the telescope for an observation override. Another useful addition to the MWA rapid-response mode is the Sun suppression subroutine in the MWA back-end scheduling web service, which places the Sun in a primary beam null during daytime triggers while also optimising the sensitivity of the target within the primary beam \citep[for the full details on this system and the GRB triggering strategy see][]{hancock19b}.

In the following we present the first rapid-response observations of a SGRB (GRB 180508A) using the upgraded MWA rapid-response mode.
In Section 2, we describe the rapid-response observations and data analysis of GRB 180805A, which pulls together different data processing and calibration techniques used for MWA transient and continuum science. 
We then describe the transient and variable search we performed using the {\sc Robbie} workflow \citep{hancock19}, followed by the image dedispersion techniques we employed to perform a dispersed transient search. 
In Section 3, we provide constraints on the prompt emission from GRB 180508A over different timescales and dispersion measures (DM), which is then followed by a discussion of the model remnant constraints that can be placed on GRB 180508A given our upper limits in Section 4. We also discuss the overall performance and potential of the MWA rapid-response mode in comparison with other low-frequency facilities with programs designed to probe for FRB-like emission associated with GRBs, and provide suggestions for a similar transient mode-of-operation for SKA-Low.

\section{MWA Observations and Data Analysis}
The MWA rapid-response observations that we present in this paper are of GRB 180805A, which was detected by \swift-BAT at 09:04:49 UT on 2018 August 5 \cite[trigger ID 851829;][]{davanzo18}. Refined analysis of the BAT light curve determined the burst had a $T_{90}$\footnote{The time interval for which 5\% to 95\% of the event's fluence was contained \citep{sakamoto11}} of $1.68 \pm 0.41$\,s \citep{palmer18} placing this GRB within the short-duration class \citep[$T_{90} \leq 2$\,s;][]{kouveliotou93}, and showed no evidence for extended emission. 
However, it is worth noting that the temporal and spectral properties of this GRB lie within the parameter space for which there may be contamination from long-duration GRBS \citep[LGRBs, produced by core-collapse supernovae; e.g.][]{galama99,stanek03} so there is a small chance this source could have a collapsar origin.

Its X-ray afterglow was subsequently detected by the \swift{} X-ray Telescope (XRT) and localised to the position $\alpha \mathrm{(J2000.0)} =  11^{\mathrm{h}}10^{\mathrm{m}}15\overset{\mathrm{s}}{.}90$ and $\delta (\mathrm{J2000.0}) = -45^{\circ}19'56\overset{''}{.}0~$ with a 90\% confidence of $2\overset{''}{.}5$ \citep{beardmore18}. 
The time-averaged spectrum from T+0.04 to T+1.90 sec is best fit by a simple
power-law model, with an index of $1.58\pm 0.32$ \citep{palmer18}. 
No other follow-up was reported.

In the following, we describe the rapid-response MWA observations of GRB 180805A, and outline the data processing pipeline and the software we employed to carry out transient searches for associated short-duration radio emission.

\subsection{MWA rapid-response observations of GRB 180805A}

The triggered Murchison Widefield Array \citep[MWA;][]{tingay13} observations of GRB 180805A were taken in the Phase II extended baseline configuration \citep{wayth18} at a central frequency of 185\,MHz (bandwidth of 30.72\,MHz) using the standard correlator mode, which has a temporal and spectral resolution of 0.5\,s and 10\,kHz, respectively. The position of GRB 180805A was observed for a total of 30\,min (15 $\times$ 2\,min snapshots). 

Each of the 15 MWA snapshot observations of GRB 180805A are summarised in Table~\ref{tab:mwa}, which includes the observation identification number (ObsID), start time, and pointing position.
As a result of the Sun avoidance code, the source's right ascension and declination were not tracked in a standard way, which meant the GRB was in a different position with respect to the primary beam for each snapshot. 
This was particularly the case between ObsIDs 1217495544 and 1217495664, where the pointing centre changed by nearly 3\,deg.
This large shift also resulted in the serendipitous placement of the bright radio source Centaurus A (Cen A) in a 10\% primary beam sidelobe for the remainder of the observation.

\begin{table*}
\begin{center}
\caption{MWA 2-minute snapshot observations of GRB 180805A}
\label{tab:mwa}
\begin{tabular}{lcS[table-format=2.2]S[table-format=3.3]S[table-format=4.3]cccc}
\\
\hline
ObsID & Start Time & {Time} & \multicolumn{2}{c}{Pointing} & Offset & RMS$^{\mathrm{~c}}$ \\
 & 2018-08-05 & {post-burst} & {RA} & {Dec} &   \\
& (UT) & {(min)} & {(deg)}  & {(deg)} & {(deg)} & (mJy)  \\
\hline
\hline
  1217495184 &  9:06:06 &   1.32   &	166.58  &  -42.715 &  2.71 & 35 \\ 
  1217495304 &  9:08:06 &   3.32   &	167.08  &  -42.715 &  2.64 & 35 \\ 
  1217495424 &  9:10:06 &   5.32   &	167.581 &  -42.715 &  2.62 & 36 \\ 
  1217495544 &  9:12:06 &   7.32   &	168.081 &  -42.714 &  2.64 & 34 \\ 
  1217495664 &  9:14:06 &   9.32   &	158.201 &  -39.878 &  8.78 & 63 \\ 
  1217495784 &  9:16:06 &  11.32   &	158.702 &  -39.878 &  8.49 & 61 \\ 
  1217495904 &  9:18:06 &  13.32   &	159.202 &  -39.877 &  8.22 & 61 \\ 
  1217496024 &  9:20:06 &  15.32   &	159.703 &  -39.877 &  7.95 & 60 \\ 
  1217496144 &  9:22:06 &  17.32   &	160.204 &  -39.877 &  7.68 & 59 \\ 
  1217496264 &  9:24:06 &  19.32   &	160.704 &  -39.876 &  7.43 & 58 \\ 
  1217496384 &  9:26:06 &  21.32   &	161.205 &  -39.876 &  7.19 & 55 \\ 
  1217496504 &  9:28:06 &  23.32   &	161.706 &  -39.876 &  6.95 & 53 \\ 
  1217496624 &  9:30:06 &  25.32   &	162.206 &  -39.876 &  6.73 & 56 \\ 
  1217496744 &  9:32:06 &  27.32   &	162.707 &  -39.875 &  6.52 & 56 \\ 
  1217496864 &  9:34:06 &  29.32   &	163.207 &  -39.875 &  6.33 & 57 \\ 
\hline
\end{tabular}
\end{center}
All raw data corresponding to the listed ObsIDs are public and can be accessed via the the MWA All-sky Virtual Observatory (\href{the Murchison Widefield Array All-sky Virtual Observatory (https://asvo.mwatelescope.org/)}{https://asvo.mwatelescope.org/}) under project code D0009. \\
$^{\mathrm{~c}}$ The RMS at the \swift-XRT position of GRB 180805A as output by \robbie{} (see Section~\ref{sec:robbie}).\\
\end{table*}

While each snapshot lasts 2\,min, the first 4\,s and last 5.5\,s of each observation are flagged.
This means that only 110.5\,s of data are collected per snapshot, which results in a loss of 9.5\,s of data between each observation (142.5\,s over the total 30\,min integration).
However, as we expect SGRBs to be at cosmological redshifts, any coherent radio signal will be highly dispersed and take tens of seconds to cross the 30.72\,MHz bandwidth at 185\,MHz (see Section~\ref{sec:late}).
Therefore the missing data cannot entirely obscure a dispersed signal, it will merely result in some loss in signal-to-noise following image-based dedispersion (technique described in Section~\ref{sec:ided}).

\subsubsection{Latencies associated with the MWA trigger of GRB 180805A}\label{sec:late}

The VOEvent associated with the \swift{}-BAT detection of GRB 180805A was circulated 64\,s post-burst (approx 18\,m before the corresponding GCN Circular was published online\footnote{\url{https://gcn.gsfc.nasa.gov/gcn3/23074.gcn3}}).
The MWA rapid-response front-end web service received the VOEvent via the Comet client 1\,s later, at which point the event was added to a queue that ensures incoming events are sent to the \swift-\fermi\ VOEvent handler in chronological order. 
The handler parses all incoming VOEvents for events of interest, and in this case captured the event associated with GRB 180805A, identifying it as a real GRB and triggering MWA observations.
The MWA was on-target and observing GRB 180805A within 20\,s of the \swift{} broadcast of the VOEvent. 
A time-line of the detection, alert, and response is summarised in Table~\ref{tab:trigger}. 
The code that queues incoming VOEvents and passes them to the various VOEvent handlers (event 3 in Table~\ref{tab:trigger}) resulted in a delay of 2.8\,s for GRB 180805A.
At this point, the front end VOEvent parser sent a request to the back-end web-service on the telescope requesting an immediate observation.
It took 1.8\,s for the back-end to calculate a pointing direction that placed the Sun in a primary beam null for the first observation (since this was a day-time observation), calculate the best calibrator to observe following the 30\,min GRB observation, and then schedule the observations.

The MWA was on target at 09:06:06 UT (77\,s post-burst), however, the correlated data recording started 2.7\,s later (this delay is related to internal MWA correlator setting changes) and the next 4\,s of data were automatically flagged by the {\sc AOFlagger} algorithm  \citep{offringa10,offringa12} in order to remove bad data from the first scan integration boundary.
However, it is possible that a subset of this 4\,s may still contain good data if using the MWA Phase II correlator, so future GRB data processing may be able to retrieve additional seconds of good data at the start of the observation.
For GRB 180805A, the very first non-flagged timestep was therefore collected by the MWA at 09:06:12.7\,UT, just 83.7\,s following the \swift{} detection.

\begin{table*}
\begin{center}
\caption{Timeline for MWA triggered observations of GRB 180805A}
\label{tab:trigger}
\begin{tabular}{lS[table-format=3.2]ll}
\\
\hline
UT Time      & {Latency} & \multicolumn{2}{l}{Event} \\
(2018-08-05) & {(s)}     & \# & Description \\
\hline
09:04:49   & 0    & 1 & \swift{}-BAT detects GRB 180805A \\
09:05:53   & 64   & 2 & \swift{} VOEvent alert notice circulated \\
09:05:54   & 65   & 3 & MWA front-end receives VOEvent via Comet and is queued for processing \\
09:05:56.8 & 67.8 & 4 & VOEvent handler processes VOEvent \\
09:05:57.1 & 68.1 & 5 & VOEvent handler sends trigger to back-end to schedule observations \\
09:05:58.9 & 69.9 & 6 & MWA schedule is updated \\
09:06:06   & 77   & 7 & MWA is on target \\
09:06:12.7 & 83.7   & 8 & MWA begins observations of SGRB 180805A \\
\hline
\end{tabular}
\end{center}
\end{table*}

\subsection{MWA data processing}\label{mwa_proc}
In the following we describe the semi-automated transient reduction pipeline used to process the triggered MWA observations of GRB 180805A (Section~\ref{sec:pipe}).
The final images output by this transient reduction pipeline are then processed using the batch processing work-flow for detection of radio transients and variables using \robbie{} \citep[][see Section~\ref{sec:robbie}]{hancock19}.
We then provide a description of the dedispersion technique we employ to search for prompt, dispersed signals \citep[e.g.][see Section~\ref{sec:ided}]{tingay15,sokolowski18}.

In order to search for coherent low-frequency radio emission associated with GRB 180805A, it is necessary to search over a wide range of cadences.
The choice of transient timescales must reflect the expected dispersion delay of the prompt radio signals for the observing frequency and bandwidth. 
Taking into account both extragalactic and Galactic ($DM=90$\,cm$^{-3}$\,pc) contributions to the dispersion measure, \citet[][see their Figure 1]{hancock19b} determined that at 185\,MHz, the arrival time of the signal could be delayed between $28-277$\,s for the observed SGRB redshift range of $0.1<z<2.5$ \citep{rowlinson13} for a line-of-sight through a Galactic latitude of $b=13.97^{\circ}$. 
In addition, they also determined it could take between $9-91$\,s for that same signal to cross the 30.72\,MHz MWA observing band at 185\,MHz.
In most cases, the MWA will likely be on target fast enough, and will integrate for long enough (usually 30\,min), to catch any prompt signal emitted during and/or several minutes following the merger (i.e. from a remnant magnetar) for most SGRBs. 
For this particular experiment, we therefore chose to create images on 5\,s, 30\,s, and  2\,min timescale for transient searches, which accommodate the length of time we might expect an associated prompt radio signal to traverse the 30.72\,MHz bandwidth for the lowest ($z\approx0.1$), average ($z\approx0.7$) and highest ($z\approx2.5$) redshifts known for SGRBs \citep{rowlinson13}. 
We also create images on 0.5\,s timescales, which is the native temporal resolution of the MWA Phase II correlator. 
Note that while we expect the prompt radio signal from a GRB to take more than 0.5\,s to cross the MWA band, we search for transients on these short timescales as a demonstration of the MWA's capability to create quality images on its native temporal resolution due to its excellent instantaneous $(u,v)$ coverage and moderate sensitivity.
Imaging on these short timescales is extremely useful for much closer transients, such as the related project to follow-up aLIGO/Virgo-detected BNS and NS-BH mergers \citep{kaplan16}. 
A mean image of the 2\,min snapshots is also created and used to search for associated persistent emission \citep[e.g.][]{totani13}. 
We also generate 24 coarse channel images (corresponding to a bandwidth of  1.28\,MHz) on each 0.5\,s timescale to facilitate a more sensitive search for dispersed, prompt, coherent signals over a wide range of dispersion measures (see Section~\ref{sec:ided}).

\subsubsection{MWA transient reduction pipeline}\label{sec:pipe}

The MWA triggered observation of GRB 180805A was processed using a semi-automated pipeline called {\sc MWA-fast-image-transients},\footnote{https://github.com/PaulHancock/MWA-fast-image-transients} which is specifically designed for the reduction of transient MWA imaging data in preparation for transient analysis.
It is therefore capable of creating images over a wide variety of timescales to match the coherent radio transient timescales expected from GRBs. 
This workflow is heavily based on the processing techniques employed for the GaLactic and Extragalactic All-sky Murchison Widefield Array (GLEAM) survey \citep{hurley-walker17}, GLEAM-X\footnote{https://github.com/nhurleywalker/GLEAM-X-pipeline/}, as well as the MWA image-plane FRB search \citep{rowlinson16}.

All observations of GRB 180805A were calibrated using the in-field calibration technique developed for the GLEAM-X pipeline. 
The in-field calibration scheme uses the GLEAM catalogue as a sky model \citep{hurley-walker17}, rather than the more traditional single bright source model. This reduces the positional shifts induced by differing ionospheric refraction in different sky locations. 
Rather than deriving in-field calibration solutions for each individual snap-shot observation of a GRB, we choose to apply a single solution derived from a snap-shot close to the centre of the 30 minute integration, which will likely have the median ionospheric shift compared to the observations on either side. The advantage of a single solution is that each calibrated observation will have a fairly consistent absolute flux calibration.
For GRB 180805A, we instead chose to use the in-field calibration solutions derived from the observation 1217495544 as the nearby bright source Cen A was in a primary beam null \citep[the final calibration solution applied to all ObsIDs can be accessed via][]{anderson21zen_mwa}. The change in the observational pointing centre that occurred between this and the following observation 1217495664 (due to the Sun suppression subroutine) resulted in Cen A sitting in a $\sim10\%$ primary beam sidelobe in all the following snapshots. At the time of processing, the sky-model used for in-field calibration did not include Cen A so if any later observations were used for the in-field calibration, the resulting solutions would include unmodelled flux.

The calibration and imaging was conducted in the following manner using scripts from the {\sc MWA-fast-image-transients} workflow (names of scripts are indicated for reproducibility).
\begin{enumerate}
\item We use {\sc obs\_asvo.sh} to fetch all target observations from the MWA All-Sky Virtual Observatory (ASVO) service\footnote{\href{https://asvo.mwatelescope.org/}{https://asvo.mwatelescope.org/}}. The ASVO service converts the MWA specific data format into a measurement set format,\footnote{https://casa.nrao.edu/casadocs/casa-5.1.0/reference-material/measurement-set/} applies phase calibration based on cable lengths and antenna/tile positions, and flags known bad tiles and channels.
\item We use an in-field calibration scheme ({\sc obs\_infield\_cal.sh}) on one of the target observations. Provided the ionosphere is reasonably stable, this would usually be the central 2\,min snapshot of the total 30\,min observation. For GRB 180805A, we used ObsID 1217495544 (see explanation above). 
During the calibration stage for GRB 180805A, we identified that tile 84 was not producing good data, so it was flagged from all observations using {\sc obs\_flag\_tiles.sh}.
\item The derived calibration solution is then applied to all of the target observations using {\sc obs\_applycal.sh}. 
\item All imaging was carried out with {\sc wsclean} \citep{offringa14}, with which we generate multi-frequency synthesis images on a range of timescales (2min, 30s, 5s, and 0.5s) using a pixel scale of 16\,arcsec and an image size of 4096 pixels ($18.2\deg$) on a side (these scales are appropriate for the Phase II extended configuration). The 2min and 30s timescale images were cleaned such that all peaks above $3\sigma$ were cleaned to $1\sigma$ ({\sc obs\_image.sh} and {\sc obs\_im28s}), and the restored images were created using a 80\,arcsec restoring beam for all observations. The 5s and 0.5s images were created using {\sc obs\_im5s.sh} and {\sc obs\_im05s.sh} but were not cleaned. All imaging scripts create instrumental Stokes XX and YY polarisation images, which are primary beam corrected and converted into Stokes I images. On 0.5s timescales, we also produce a set of images that are divided into 24 coarse channels of bandwidth 1.28\,MHz using {\sc obs\_im05s\_24c.sh}.
\end{enumerate}

All the output images are then prepared for transient and variable analysis using {\sc Robbie} (see Section~\ref{sec:robbie}) as this work-flow requires that all images have the same shape and sky coverage.
As each of the 15 snapshot observations has a different pointing centre, all images were re-gridded to a template image that represents the largest region of overlap between snapshots (a nearly square region $\sim140$\,deg$^{2}$ in size centred at $\alpha \mathrm{(J2000.0)} = 10^{\mathrm{h}}48^{\mathrm{m}}55\overset{\mathrm{s}}{.}7$ and $\delta (\mathrm{J2000.0}) = -40^{\circ}27'50\overset{''}{.}4$) using the {\sc Miriad} \citep{sault95} task {\sc regrid}.

\subsubsection{Transient and variable search}\label{sec:robbie}

We used \robbie{} to search for transient and variable sources on all full-bandwidth (30.72\,MHz) images, grouped by cadence (2\,min, 30\,s, 5\,s, and 0.5\,s). 
The scale of data being processed for this project was beyond the capability of \robbie{} as described by \citet[`V1.0'][]{hancock19}. 
The main limitation was the use of {\sc Make} \citep{feldman_make_1979} as a workflow manager
in an HPC environment, as it does not integrate with queue scheduling systems such as SLURM \citep{yoo_slurm_2003}.
Additionally, V1.0 was developed and tested with 25 epochs of data, however when scaling out to the 3315 epochs required for the 0.5\,s images in this project, it became clear that the processing and storage choices were inappropriate. In particular, FITS\footnote{\url{https://fits.gsfc.nasa.gov/fits\_standard.html}} tables are limited to 1000 columns.
To counter these limitations, we ported \robbie{} to use {\sc NextFlow} \citep{DiTommaso2017} as the workflow manager, containerised the entire software stack using Docker\footnote{\url{https://www.docker.com/}} and Singularity \citep{kurtzer_singularity_2017}, and used VOTables\footnote{\url{https://www.ivoa.net/documents/VOTable/20191021/}} instead of FITS files to store tabulated data. This updated version of \robbie{} is now more portable, can be run on desktop or HPC systems, and can scale to surveys with thousands of epochs of observations. The logical flow and operation of \robbie{} is unchanged and we summarise its operation below.

\robbie{} first derives and then applies an astrometric correction to each image that corrects for any ionosphere distortion.
This is conducted using the algorithm {\sc fits\_warp} \citep{hurley-walker18}, and uses the GLEAM catalogue as a template for the source positions (note that due to the minimal signal detected in the coarse channel images, the ionospheric distortions calculated for the full-bandwidth 0.5\,s images were applied to the 24 coarse channel images of the corresponding timestep, which were used in the analysis described in Section~\ref{sec:ided}).
Once the ionospheric corrections are applied, \robbie{} proceeds with a transient and variable search as follows:

\begin{itemize}
    \item Stack all epochs to form a cube;
    \item Create a mean and noise image from the cube;
    \item Use \aegean{} blind source finding to identify persistent sources in the mean image;
    \item Use \aegean{} priorized fitting to create a light curve for each persistent source;
    \item Mask persistent sources from each individual epoch; and
    \item Use \aegean{} blind source finding on the masked single epoch images to identify transient candidates (which may appear in one or more epochs).
\end{itemize}

\robbie{} was modified to allow positions of interest to be added to the persistent source catalogue, causing light curves to be generated for these locations via priorized fitting.
This functionality was used to calculate a light curve at the \swift{}-XRT position of \grb{} (see Section~\ref{sec:lcresults}). 

Corrections to the absolute flux scale were calculated using the software tool {\sc flux\_warp} \citep{duchesne20}, which compares the source fluxes measured by \aegean{} to the GLEAM catalogue.
We assumed a mean offset in flux for each of the 2\,min snapshot images and applied this mean correction to all quoted flux densities and limits in Table~\ref{tab:mwa}, Table~\ref{tab:rms}, and Section~\ref{sec:results}. 
Deriving reliable absolute flux-scale corrections on shorter timescales was not feasible due to fewer sources being detected, and the source flux density uncertainties being larger.
We therefore assumed that the flux calibration did not change within a 2\,min snapshot and applied the same correction to the shorter timescales (30\,s, 5\,s and 0.5\,s).
The flux scale corrected light curves at each timescale are shown in Figure~\ref{fig:lc_monitoring}. 
Following the mean correction on the 2\,min snapshot images, we measured a $1\sigma$ scatter on the final flux values of $\leq15$\%.

\subsubsection{Image de-dispersion and dispersed transient search}\label{sec:ided}

Searching for dispersed transient signals via image de-dispersion techniques is potentially far more sensitive to these likely narrow (ms timescales) FRB-like signals when compared to image searches for which the integration time is equivalent to how long we expect the signal to traverse the full 30.72\,MHz bandwidth for a given DM \citep[e.g.][]{REALFAST2018}.
For example, if the FRB has an intrinsic duration of 100\,ms at a redshift of 0.7, then it will take 30\,s to cross the MWA bandwidth so a snapshot of this duration would result in a signal-to-noise loss of a factor of 300.
The final stage of our data processing for GRB 180805A is to perform a search for dispersed, prompt signals on the native MWA correlator temporal resolution of 0.5\,s and the coarse channel resolution of 1.28\,MHz.
The recent detection of FRB~191001, which was detected in an ASKAP snapshot image with a 10\,s integration time and later processed with a DM search, demonstrates the utility of a dual processing approach \citep{Bhandari2020}.
This search was conducted over the full 30\,min observation for a DM range of $200-3000$\,pc\,cm$^{-3}$, which corresponds to the known SGRB redshift range of $\sim0.1<z<2.5$.
We compute the dispersive delay of a pulse sweep between the upper ($\nu_h$) and lower ($\nu_l$) ends of a band for a given DM as:
\begin{eqnarray}
   \delta t (DM,\nu_l,\nu_h)& = & 4.15~{\rm DM}\,\left(\frac{1}{\nu^2_l} - \frac{1}{\nu^2_h} \right)~[{\rm ms}], \label{eq_dispersion}
\end{eqnarray}
where frequencies $\nu_l$ and $\nu_h$ are expressed in GHz. The MWA band for the observations of GRB 180805A is between $\nu_l = 0.175$\,GHz and $\nu_h = 0.200$\,GHz.  

This analysis was performed on the 0.5\,s duration coarse channel images that were output by the {\sc MWA-fast-image-transients} pipeline (see Section~\ref{mwa_proc} point iv).
The ionospheric correction that \robbie{} calculated for each 0.5\,s image (see Section~\ref{sec:robbie}) was applied to each set of coarse channel images (of which there are 24) for the corresponding timestep.
It was not possible to calculate the ionospheric correction on individual 0.5\,s coarse channel images due to their low sensitivity preventing the detection of individual field sources. 
Ionospheric induced position shifts are frequency dependent ($\propto 1/\nu^2$), with larger offset distortions expected at lower frequencies.
The expected difference in position shifts between the upper and lower end of the observing band are just $(\Delta\nu/\nu)^2\sim3\%$ for a 30.72\,MHz bandwidth at 185\,MHz.
We did not apply a chromatic correction to the ionospheric offsets calculated from the 0.5\,s images over the full 30.72\,MHz bandwidth before applying them to the coarse channel images.

\begin{figure*}
  \begin{center}
  \includegraphics[width=\textwidth]{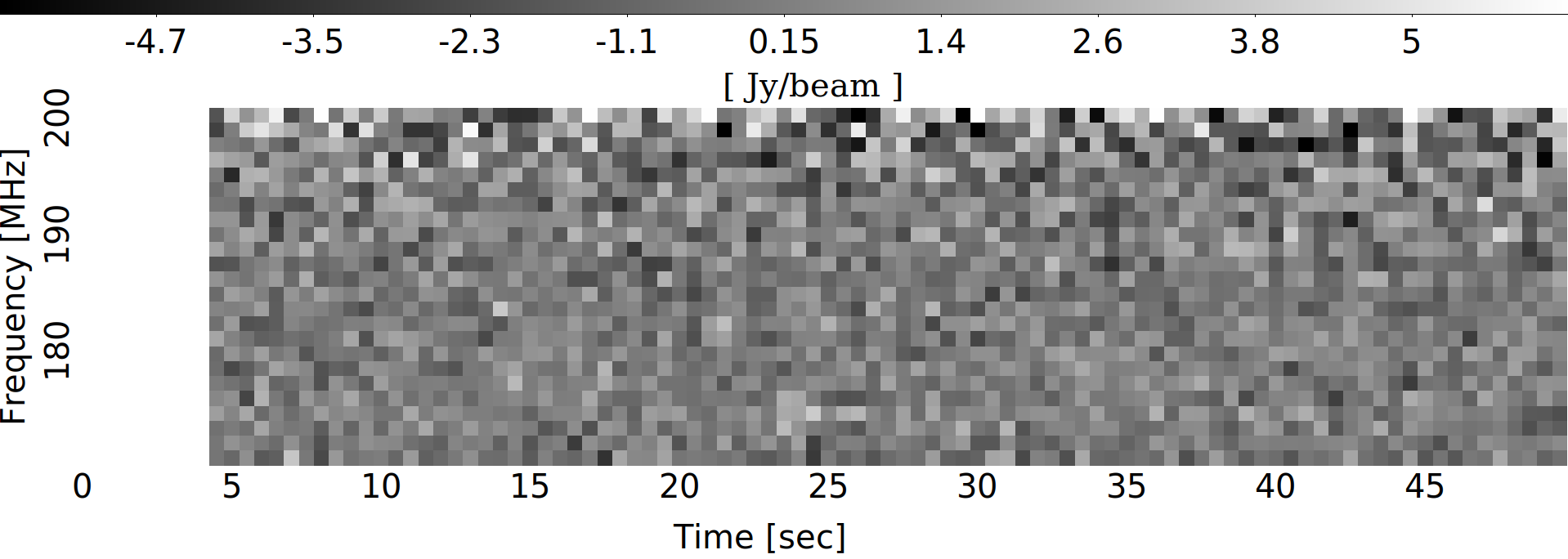} 
  \caption{The first 100\,s of the dynamic spectrum measured at the pixel location of GRB\,180805A at a 1.28\,MHz frequency and 0.5\,s time resolution. The full MWA follow-up observation covered approximately 30\,min. The first 4\,s and last 5.5\,s of each individual observation are flagged due to instrumental reasons. While the first timestep begins at $t=$79.68\,s post-burst, the first 4\,s are flagged so the first non-flagged integration starts at 09:06:12.7 UTC ($83.68$\,s post-burst). The full dynamic spectrum can be downloaded from \citet{anderson21zen_mwa}.}
  \label{fig_dynaspec}
  \end{center}
\end{figure*}

The 0.5\,s coarse channel images were first used to create a $n_{t} \times n_{ch}$ dynamic spectrum at the position of the GRB over the full 30 minute observation. As the \swift{}-XRT $3\sigma$ positional uncertainty for GRB 180805A was 4.7\,arcsec, which is far less than the MWA Phase II synthesised beam of $\sim80$\,arcsec at 185\,MHz, it was therefore sufficient to perform our signal search analysis on a single pixel (sized 16\,arcsec square) at the nominal position of the GRB. 
A small section of this dynamic spectrum showing the first 100\,s of the MWA follow-up observation at the pixel location of GRB 180805A is shown in Figure~\ref{fig_dynaspec}. Note that the first timestep (when the MWA was on-target) began at $t=$76.98\,s post-burst, but the first 2\,s are not recorded and a further 4\,s are always flagged due to instrumental reasons. 
Therefore, the first non-flagged integration started at 09:06:12.7 UTC  ($83.68$\,s post-burst). 
The total number of time-steps is $n_{t}=3434$ including the 9.5\,s gaps between each 2\,min snapshot. 

In order to determine the noise in the dynamic spectrum in the direction of the GRB, we first chose independent directions on the sky and created additional dynamic spectra, the spectrum at the GRB location is labelled as the `signal' while the remainder are labeled as `noise'.
In order to remove instrumental effects and persistent astrophysical sources from each dynamic spectrum, we measured a time-averaged spectral energy distribution (SED), and subtracted this from each time step.
This SED subtraction was done independently for all pointing directions.
The dynamic spectral plots for the `noise' directions were combined into a cube, which were then used to compute a mean ($\mu$) map. 
Finally, we computed a final GRB dynamic spectrum (DS) by subtracting the mean from the signal:
\begin{equation}
    \mathrm{DS} = \mathrm{signal} -\mu
\end{equation}

Using the noise subtracted GRB dynamic spectrum, a de-dispersed time series (DTS) was calculated for each DM in the range [$D_{min},D_{max}]$ and potential pulse start time ($T_{s}$) in the range ($[-\tau_{max} , T_{last}]$), where $D_{min}$ is the minimum trialed DM
(200\,$\mathrm{pc\,cm^{-3}}$), $T=0$\,s is the start time of the first recorded MWA timestep, $\tau_{max} \approx122.4$\,s is the maximum sweep time over the MWA band ($30.72$\,MHz) corresponding to the maximum DM ($D_{max} = 3000$\,pc\,cm$^{-3}$), and $T_{last}$ is the time of the last recorded MWA timestep ($3434 \times 0.5$\,s in this case). Some of the FRBs show significant temporal pulse broadening due to scattering \citep{2020MNRAS.497.1382Q}. These scattering times at GHz frequencies correspond to a few seconds at the MWA frequency assuming frequency scaling proportional to $\sim \nu^{-4}$\,\citep{2004ApJ...605..759B}. Therefore, we have also performed averaging of dynamic spectrum along the time axis on timescales between 1 to 10\,s and then formed DTS using the time averaged dynamic spectra. Similarly, no statistically significant candidates were found.

An SGRB with a high DM could potentially allow observations at MWA observing frequencies to detect prompt radio signals that are predicted to be emitted prior to the merger (see Section~\ref{sec:coherent}).
The GRB was detected by \swift{}-BAT just 83.7\,s before the first MWA data were collected. Based on Equation~\ref{eq_dispersion}, this means that for a prompt radio signal emitted at the time of the SGRB (which we assume corresponds to the first detection of $\gamma$-ray emission) to be detectable in our MWA observing band (170--200\,MHz) we required a $DM\gtrsim800$\,pc\,cm$^{-3}$.

Therefore, for the higher DM trial values (in the range $800 - 3000$\,pc\,cm$^{-3}$) we can use the full observing band sensitivity to probe the parameter space in which radio prompt signals were emitted from the SGRB up to about 350\,s before the merger (first detection of $\gamma$-rays). 
For the DM range $580 - 800$\,pc\,cm$^{-3}$, we can use part of the observing band to search for dispersed signals emitted at the time of the SGRB as according to Equation~\ref{eq_dispersion},  $DM=580$\,pc\,cm$^{-3}$ corresponds to a 83\,s dispersive delay between the first $\gamma$-ray detection and the lower end of the MWA observing band (170\,MHz). 

\begin{figure*}
  \begin{center}
  \includegraphics[width=0.9\textwidth]{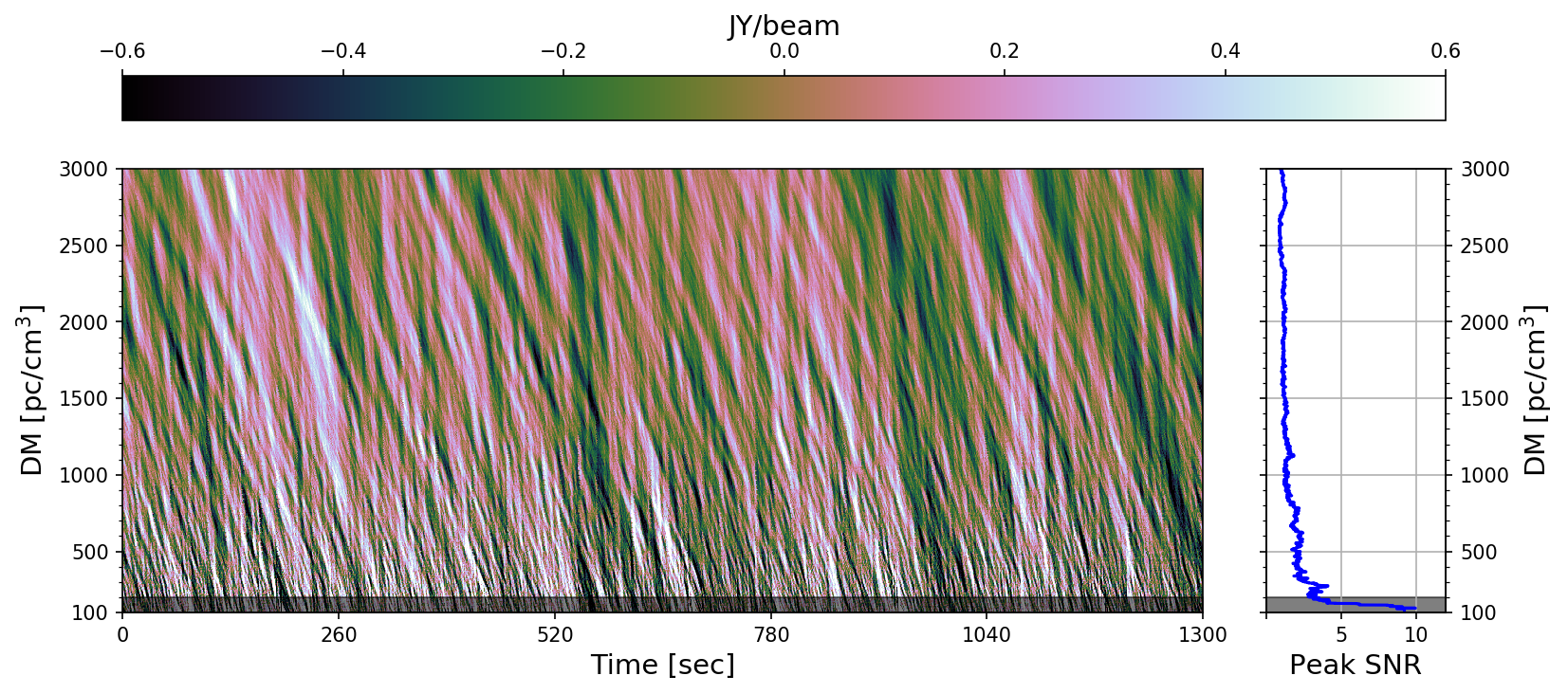}
  \caption{Left: The central region of the de-dispersed time series showing the mean flux density for the path swept across the dynamic spectrum (Figure~\ref{fig_dynaspec}) for every DM trial ranging from 100 to 3000 $\mathrm{pc\,cm^{-3}}$ at 1.0 $\mathrm{pc\,cm^{-3}}$ intervals, covering approximately $700 - 1300$\,s at a resolution of 0.1\,s following the GRB detection. Right: The peak SNR over all de-dispersed time series as a function of DM trial. The highest SNR values detected in the de-dispersed time series (up to $\approx 10$) correspond to low DMs ($<$140 $\mathrm{pc/cm^{3}}$), where the number of averaged pixels in the dispersion sweep is low and a single channel RFI can increase the de-dispersed value. Therefore, we have excluded these low DMs from our analysis. Also note that we do not expect SGRBs to have a $\mathrm{DM}<200\mathrm{pc\,cm^{-3}}$ (shaded areas) if we assume a minimum SGRB redshift of $z \sim0.1$. The full de-dispersed time series can be downloaded from \citet{anderson21zen_mwa}.
  }
  \label{fig_dts}
  \end{center}
\end{figure*}

The resulting DTS is a two dimensional image with the horizontal axis corresponding to the tested pulse start time ($T_{s}$) with a 0.1\,s resolution and the vertical axis corresponding to the DM in steps of 1\,pc\,cm$^{-3}$. 
The sampling of pulse arrival times (0.1\,s) was chosen to be smaller than the actual temporal resolution of the images to ensure no signal was missed due to rounding errors caused when calculating the positions of the pixels along the dispersion sweeps. 
Following equation~\ref{eq_dispersion}, we calculated the value in each pixel at ($T_{s}$,DM) as the mean of all pixel values (in $\JyPerBeam{}$) along the corresponding tested path swept out by a pulse across the dynamic spectrum between 170 and 200\,MHz. 
The central time region of the DTS image calculated for the central pixel (nominal position of GRB 180805A) is shown in the left panel of Figure~\ref{fig_dts}.

The standard deviation of the DTS is not constant across the image as it depends on the number of pixels averaged together for a given DM sweep, which increases with increasing DM as the dispersion sweep across the dynamic spectrum becomes longer before it reaches the lower frequency end. 
We therefore generated a corresponding map of the number of pixels ($n_{d}$) used to calculate each image pixel in the DTS image shown in the left panel of Figure~\ref{fig_dts}. 
This resulting map was used to derive the dependence of the DTS standard deviation ($\sigma_{dts}$) as a function of $n_{d}$, such that $\sigma_{dts}(n_{d}) = \sigma_{dts}(0) / \sqrt{n_{d}} $, which was divided through the DTS image to create a DTS image in units of SNR. 
Each pixel ($T_{s}$,DM) of the final SNR DTS image with a DM between 200-3000\,pc\,cm$^{-3}$ was then searched for any dispersed signals above the specified thresholds of $\sigma_{dts}>5$ and $\sigma_{dts}>6$, for which Gaussian statistics predict there to be approximately 2.76 and 0.01 random events (when only different dispersion sweep paths through the dynamic spectrum are taken into account), respectively, based on the number of DM and time trials. We verified that values in the DTS images have a Gaussian distribution but we also note that those 
dispersion sweeps with 1 or 2 pixels in common are not fully statistically independent. However, this is only a few percent effect given that even the lowest trail $\mathrm{DM} = 200\mathrm{pc\,cm^{-3}}$ has $\gtrsim50$ pixels.

We used the DTS SNR image of the GRB to further explore how the summed number of pixels effects the final SNR measurements for a given DM by plotting the peak SNR in each de-dispersed time series as a function of DM trial (i.e. the maximum SNR value for each DM over all possible signal start times; see the right panel of Figure~\ref{fig_dts}). 
We found that the high SNR ($\lesssim10$) dispersed transient candidates for GRB 180805A were grouped at $\mathrm{DMs}\leq140$\,pc\,cm$^{-3}$. 
For this low DM region, the signals were calculated by averaging a low number of dynamic spectrum pixels ($n_{d}$) due to the short dispersion sweep, with the pixel count often being further reduced due to the sweep encountering 9.5\,s gaps in the data between two MWA observations.
In the case of a single high value pixel in the dynamic spectrum (e.g. caused by RFI), this could significantly increase the mean value of a small number of pixels averaged over a short dispersion sweep, therefore boosting the final de-dispersed SNR for the corresponding trial. 
However, such an anomaly does not impact our analysis as we expect SGRBs to lie within the DM range of 200--3000\,pc\,cm$^{-3}$. 
Indeed, assuming an extra-galactic origin for GRB 180805A, we may expect a Galactic DM contribution of $\approx178$\,pc\,cm$^{-3}$ estimated from \citet{2019ascl.soft08022Y} \citep[$\approx$ 135 pc\,cm$^{-3}$ from the NE2001 model from ][]{2002astro.ph..7156C} along this line of sight.
Additionally, contributions from the MW halo and host galaxy to the DMs of localized FRBs have been found to be on average $75$ pc\,cm$^{-3}$ \citep{Macquart2020}, which is  consistent with the lowest DMs of observed FRBs: 180729.J1316+55 \citep{CHIME2019a} and FRB~171020 \citep{shannon18nat}, with excess DMs of 78.4 pc\,cm$^{-3}$ and
75.7 pc\,cm$^{-3}$, respectively.
We could therefore expect a minimum DM of $210$ pc\,cm$^{-3}$ for GRB 180805A.

\begin{figure*}
  \begin{center}
  \includegraphics[width=0.8\textwidth]{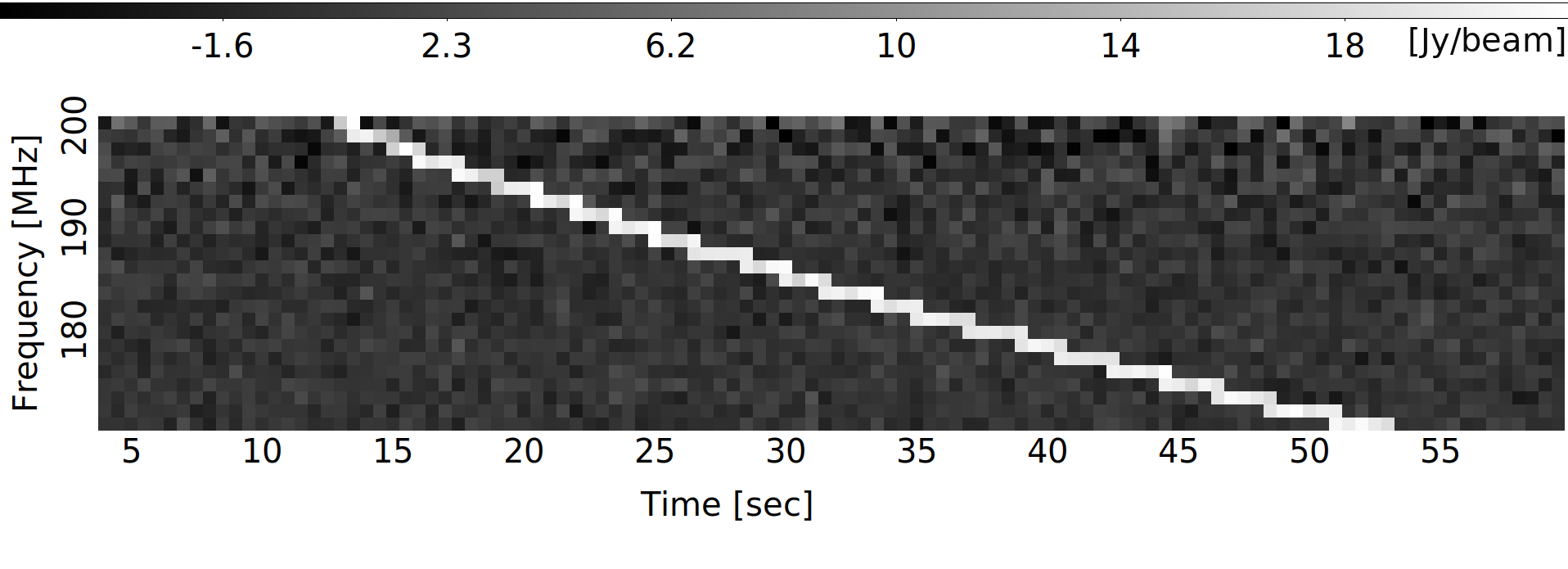} 
  \includegraphics[width=1.3in]{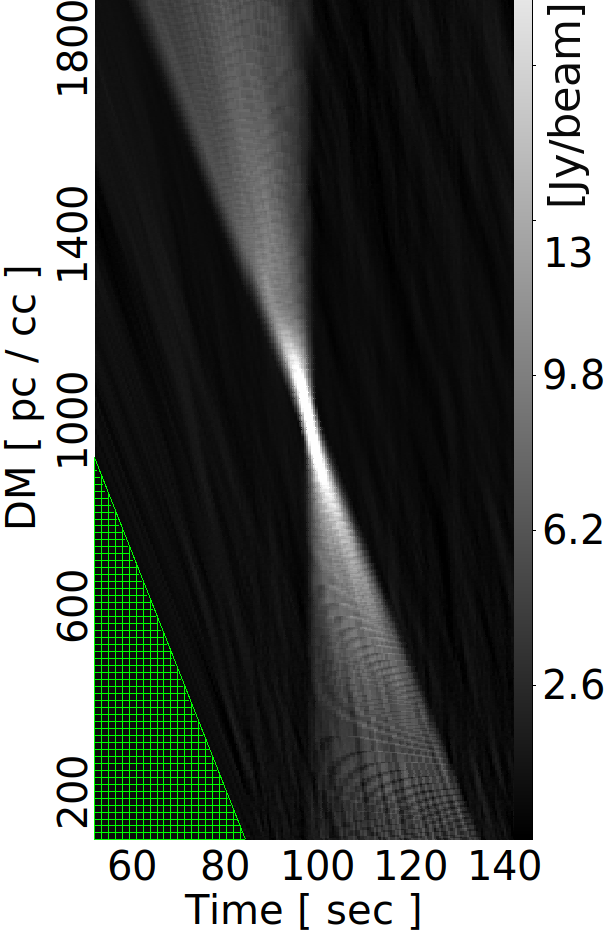}
  \caption{Left: An example of a simulated dispersed pulse with fluence 10000\,Jy\,ms and DM = 1000 $\mathrm{pc\,cm^{-3}}$ added to the dynamic spectrum of an image pixel used for this efficiency test. Again, the dynamic spectrum begins at the first MWA timestamp of 79.68\,s but the first 4\,s are flagged as shown in Figure~\ref{fig_dynaspec}. Right: The corresponding de-dispersed time series (DTS) around the start time and DM of the injected pulse. The horizonal axis start time in this DTS image corresponds to the \swift{} detection time of GRB 180805A, which was 83.7\,s before the MWA began observing the event.
  This algorithm therefore probes possible pulse start times that began before the start of the MWA follow-up observations if emitted at a high enough DM for the signal to be dispersed by $>83.7$\,s. 
  The ``green-hatched triangle'' in the bottom-left corner corresponds to negative pulse start times (radio signals arriving at the upper end of the frequency band before the MWA started observations) and DM sweep times too short to be captured at the lower frequency end ($T_s + \delta t (DM,0.170,0.200) < 0$), hence covering the DM/pulse start time parameter space inaccessible to the analysis. 
  As 100 pulses were generated for each pair of DM (12 values) and fluence (21 values), a total of 25200 simulated pulses were processed through the transient search pipeline used to search for signals in the GRB DTS described in 
  Section~\ref{sec:ided},  allowing us to determine the efficiency of this technique (see also Figure~\ref{fig_eff}). }
  \label{fig_example_injected_pulse}
  \end{center}
\end{figure*}

\begin{figure*}
  \begin{center}
  \includegraphics[width=0.95\textwidth]{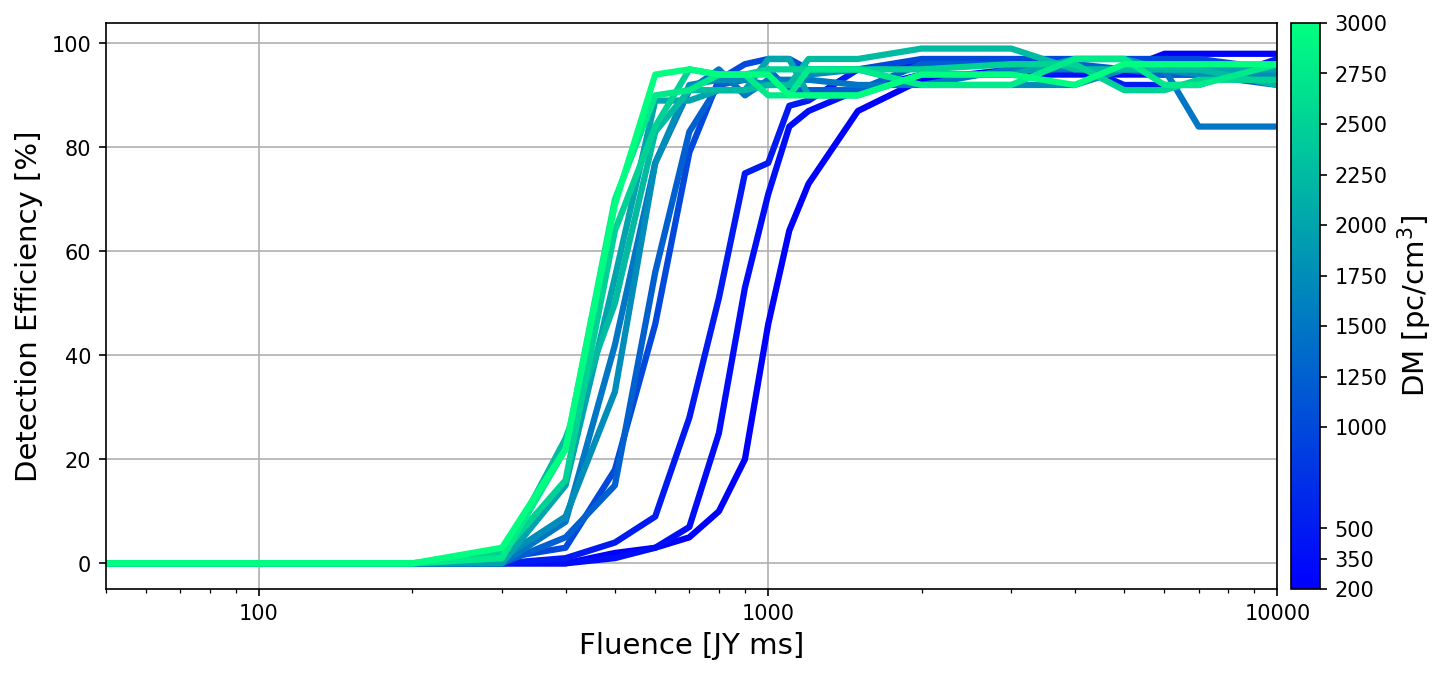}
  \caption{The measured efficiency for detecting pulses with $\sigma_{dts}> 6$
  as a function of fluence at twelve tested DMs (values indicated on the colour bar). The fluence where detection efficiency exceeds a threshold of 90\% varies as a function of DM from $\approx$1750\,Jy\,ms at DM=$200$\,$\mathrm{pc\,cm^{-3}}$ to $\approx$570\,Jy\,ms at DM=$3000$\,$\mathrm{pc\,cm^{-3}}$. Hence, any dispersed radio signal from the GRB\,180805A brighter than 1750\,Jy\,ms in the trialled DM range $200 - 3000$\,$\mathrm{pc\,cm^{-3}}$ should be detected  with 90\% probability by the search procedure described in Section~\ref{sec:ided}.
  }
  
  \label{fig_eff}
  \end{center}
\end{figure*}

\begin{figure*}
  \begin{center}
  \includegraphics[width=0.95\textwidth]{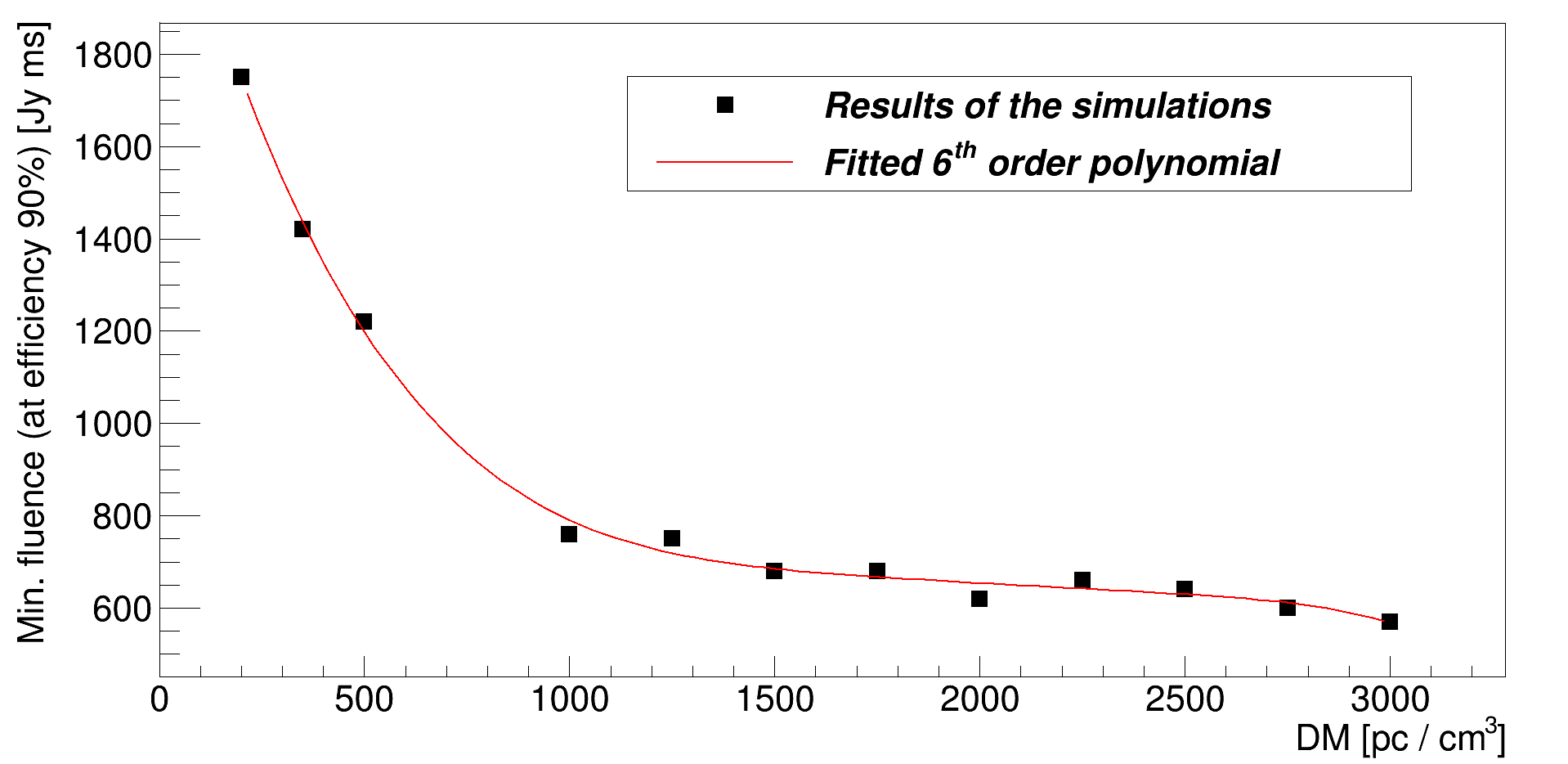}
  \caption{The 90\% fluence detection efficiency as a function of DM. A 6-order polynomial has been fitted in order to interpolate the 90\% fluence detection efficiency for any DM value within the tested range of $200-3000$\,pc\,cm$^{-3}$.}
  \label{fig_eff_dm}
  \end{center}
\end{figure*}

This de-dispersed transient search procedure was verified by simulating dispersed pulses and adding their fluxes to the appropriate pixels of a dynamic spectrum based on the dispersion sweep calculated using equation~\ref{eq_dispersion}.
We randomly simulated 100 pulses per DM and fluence pair (DM,$\mathcal{F}$) for twelve DM values (200, 350, 500, 1000, 1250, 1500, 1750, 2000, 2250, 2500, 2750 and 3000 $\mathrm{pc\,cm^{-3}}$) and twenty one $\mathcal{F}$ values (50, 100, 200, 300, 400, 500, 600, 700, 800, 900, 1000, 1100, 1200, 1500, 2000, 3000, 4000, 5000, 6000, 7000 and 10000 [Jy\,ms]). 
Currently, the algorithm only simulates narrow pulses, such that the entire fluence is contained within a single 0.5\,s time bin. 
However, simulating temporally broadened pulses due to scattering will be implemented in a future version. 
The arrival times of the simulated pulses at the upper end of the MWA observing band were randomised within the 30\,min of the MWA observations. An example of an injected pulse is shown in Figure~\ref{fig_example_injected_pulse}.
The dynamic spectra with injected pulses were processed using the same analysis as was used to search for pulsed signals in the GRB DTS (Figure~\ref{fig_dts}). 
The efficiency ($\epsilon$) of the search was calculated as $\epsilon = N_{det} / N_{gen}$, where $N_{det}$ is the number of simulated pulses identified by the algorithm and $N_{gen}$ (100 in this simulation) is the total number of generated pulses for a given (DM,$\mathcal{F}$) pair.
Figure~\ref{fig_eff} shows the pulse detection efficiency (at $\ge 6\sigma_{dts}$) for each tested DM and fluence pair. 
This demonstrates that for a threshold of $\sigma_{dts} > 6$, the 90\% fluence detection efficiency limit varies as a function of DM within the tested range, which ranges between 570\,Jy\,ms at DM$=3000$\,pc\,cm$^{-3}$ and 1750\,Jy\,ms at DM=$200$\,pc\,cm$^{-3}$, with likely zero false positives due to noise events. 
To illustrate this dependence, we performed polynomial fits to the DM curves in Figure~\ref{fig_eff} to find the 90\% fluence detection efficiency for each tested DM, which are plotted in Figure~\ref{fig_eff_dm}. The variation in efficiency is mainly due to the noise in the dynamic spectrum.

\section{Results}\label{sec:results}

\subsection{Light curve analysis}\label{sec:lcresults}

\begin{figure}
    \centering
    \includegraphics[width=\linewidth]{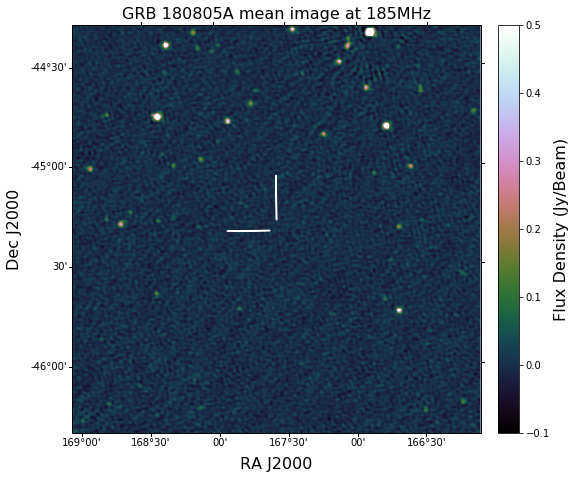}
    \caption{Mean (averaged) image of the MWA 30\,min integration (2\,min $\times$ 15 snapshot observations) of GRB 180805A, beginning 83.7\,s post-burst. The image size is $2^{\circ} \times 2^{\circ}$ centered on the XRT position of GRB 180805A ($2\overset{''}{.}5$ 90\% confidence), which is also indicated by two white lines that point to within $3'$ of this position. The image RMS at the position of GRB 180805A is 13.4\,mJy.}
    \label{fig:image}
\end{figure}

The \robbie{} workflow pipeline was used to perform a blind search for transients within the field of GRB 180805A over multiple timescales (2\,min, 30\,s, 5\,s, and 0.5\,s), however, none were detected. A summary of the RMS at the position of GRB 180805A can be found in Table~\ref{tab:mwa} for each of the 2\,min snapshots, and the full range in RMS at the position of GRB 180805A on the four different timescales are also summarised in Table~\ref{tab:rms}. The mean image output by \robbie{} that was created from the 2\,min snapshots (equivalent to the 30\,min integration) is shown in Figure~\ref{fig:image}, where the \swift{}-XRT position of GRB 180805A is indicated. No associated persistent radio source was detected with a $3\sigma$ upper limit of 40.2\,mJy.

In order to better evaluate any potential variability properties of GRB 180805A, we use \aegean{} priorized (forced beam) fitting in the \robbie{} workflow to measure the flux density at the \swift{}-XRT position over all four timescales, producing light curves that are shown in Figure~\ref{fig:lc_monitoring} \citep[these light curve data can be accessed via][]{anderson21zen_mwa}.
The RMS (and three times the RMS) at the GRB position are also shown.
The jump in the RMS noise at $\sim600$\,s post-burst (ObsID 1217495664) is due to the large shift in the observational pointing centre caused by the Sun avoidance algorithm, which means GRB 180805A was sitting in a less sensitive part of the primary beam from this ObsID onward.
Another contribution to the noise may also be caused by Cen A, which is not included in the GLEAM model and was sitting in a 10\% primary beam sidelobe for the same ObsID range.

\subsubsection{Variability analysis}
\label{sec:variability}

\begin{figure}
    \centering
    \includegraphics[width=1.2\linewidth]{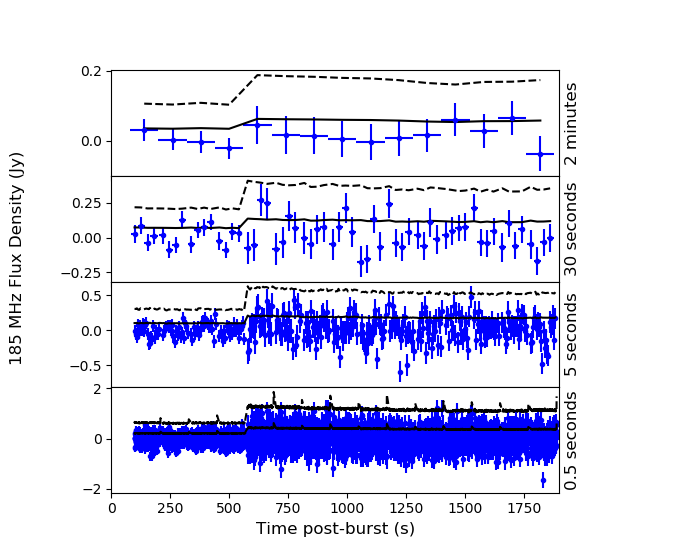}
    \caption{Monitoring light curves at the position of GRB 180805A over four different timescales (2\,min, 30\,s, 5\,s, and 0.5\,s) as a function of time following the \swift{}-BAT detection. The plotted flux densities (blue data points) were measured via performing priorized fitting at the \swift{}-XRT position of the GRB using \robbie{}. All flux densities are plotted at the centre of each time-bin. The vertical errorbars correspond to the flux density error output by \robbie{} and the horizontal errorbars indicate the duration of the time-bin. The solid black line shows the RMS for each time-bin at the position of the GRB, with the $3\sigma$ level (3 $\times$ RMS) indicated by the dashed black line. Statistics quoted in Table~\ref{tab:rms} indicate no significant variability was observed at the position of GRB 180805A on any of the four timescales.
    The csv files listing the resulting flux density as a function of time for these four timescales can be downloaded from \citet{anderson21zen_mwa}. 
    }
    \label{fig:lc_monitoring}
\end{figure}

The {\sc Robbie} workflow calculates variability statistics for each light curve it generates. In Table~\ref{tab:rms}, we quote the modulation index ($m = \frac{\sigma}{\mu}$), the de-biased modulation index 
\citep[$m_d$; which takes into account the errors on each data-point, see][]{hancock19}, 
and the probability of observing such variability in a non-variable (steady) source (p\_val). 
Initially {\sc Robbie} was designed with the assumption that a light curve would have fairly consistent uncertainties on all observations such that the p\_val parameter could be computed from a $\chi^2$ statistic and an effective number of degrees of freedom.
However, when the telescope repoints part way through the observation, the instrumental sensitivity changes (see Figure\,\ref{fig:lc_monitoring}).
For GRB\,180805A, the light curve of a non-variable source is therefore no longer drawn from a normal distribution, but a population that is the sum of two normal distributions with a common mean but different variance.
We have therefore modified the p\_val calculation performed by {\sc Robbie} so that it will first normalise the light curve to a zero mean and unit variance via:
\begin{equation}
    Z_i = \frac{x_i - \mu}{\sigma_i}
\end{equation}
where $x_i$ and $\sigma_i$ are the measurement and associated uncertainty, and $\mu$ is the mean flux.
The modified p\_val is now computed by performing a Kolmogorov-Smirnov test \citep{smirnov_estimation_1939} of $Z$ against a normal distribution using the {\sc scipy.stats.kstest} python function.
The resulting p\_val quoted in Table~\ref{tab:rms} show there is no significant variability on any of the four timescales at the position of GRB 180805A. 
For details on how {\sc Robbie} calculates the remaining statistics, see \citet{hancock19}. 

\begin{table*}
\begin{center}
\caption{Light curve variability statistics and range in the RMS at the position of GRB 180805A for different monitoring timescales. Quantities reported are the modulation index ($m$), the debiased modulation index ($m_d$), and the probability that the this variability will be observed in a non-variable source (p\_val). See \S{\ref{sec:variability}} for a description of these metrics.
}
\label{tab:rms}
\begin{tabular}{S[table-format=3.2]S[table-format=3.2]S[table-format=3.2]S[table-format=3.2]c}
\\
\hline
{Timescale} & \multicolumn{3}{c}{Variability statistics} & RMS \\
{(s)} & $m$ & {$m_d$} & p\_val & (mJy beam$^{-1}$)  \\
\hline
0.5 & 18.44 & 6.34 & 0.19 & $190-610$ \\ 
5 & 8.54 & 3.59 & 0.85 & $90-210$ \\ 
30 & 4.74 & 1.03 & 0.4 & $65-140$ \\ 
120 & 1.84 & -2.59 & 0.86 & $35-65$ \\ 
\hline
\end{tabular}
\end{center}
\end{table*}

\subsection{Image de-dispersion analysis}\label{sec:results_ided}

As described in Section~\ref{sec:ided}, the final DTS SNR image of the pixel position of GRB 180805A was searched for trials ($T_{s}$,DM) for signals exceeding thresholds of $5\sigma_{dts}$ and $6\sigma_{dts}$. 
No ($T_{s}$,DM) trials with $\sigma_{dts} \geq 6$ were identified for the  DM range $200-3000\,\mathrm{pc\,cm^{-3}}$, while two $\sigma_{dts} \geq 5$ trials were identified, which is in agreement with the statistical expectations of the false-alarm-rate (see Section~\ref{sec:ided}). 
As a sanity check, all the SNR$>5$ candidates were ``eye-balled'' and none of them had a corresponding FRB-like signal dispersion sweep in the dynamic spectrum. 
We conclude that we did not detect any SNR$>6$ prompt radio signals associated with the short GRB\,180805A in the de-dispersed MWA data in the DM range corresponding to the observed SGRB redshift range, with fluence upper-limits ranging between 570\,Jy\,ms at DM$=3000$\,pc\,cm$^{-3}$ and 1750\,Jy\,ms at DM=$200$\,pc\,cm$^{-3}$.
The dynamic spectrum at the pixel position of GRB 180805A and the final DTS can be accessed via \citet{anderson21zen_mwa}.

\section{Discussion}
The MWA rapid-response follow-up observations of GRB 180805A have allowed us to probe for prompt, coherent, dispersed radio signals that may have been emitted either just prior to or shortly following the merger event.
No transient signals were detected at the position of GRB 180805A over a variety of timescales that cover the spread of dispersion measures expected for the minimum, average and maximum redshift range of SGRBs. However, we have obtained some of the most 
stringent limits on short timescales (0.5\,s, 5\,s, 30\,s and 2\,mins starting just 83.7\,s post-burst) at low radio frequencies ($<300$\,MHz) for an SGRB (see Table~\ref{tab:rms}).

Our image plane dedispersion analysis demonstrated that the rapid-response MWA observation of GRB 180805A was sensitive to dispersed, prompt radio signals over a range of fluences as a function of DM (570\,Jy\,ms at DM$=3000$\,pc\,cm$^{-3}$ up to 1750\,Jy\,ms for DM$=200$\,pc\,cm$^{-3}$; see Figures~\ref{fig_eff} and \ref{fig_eff_dm}).
The more stringent fluence limits at higher DMs are due to the signal search being conducted over many more dynamic spectrum pixels than at lower DMs.
The range of fluence limits derived in our analysis are high in comparison to the (higher frequency) Parkes and ASKAP FRB population \citep{petroff16}\footnote{\url{http://www.frbcat.org/}}, with the brightest known event being FRB 180110 at $420 \pm 20$\,Jy\,ms \citep{shannon18nat}.
\citet{james19a} assumed a power-law distribution for the rate $(R)$ of FRBs with a fluence threshold of $(\cal F)$ according to

\begin{equation}\label{eq:frb}
    R({\cal F}) = R_{0} \left( \frac{\cal F}{{\cal F}_0} \right)^{\alpha} \mathrm{sky^{-1}\,d^{-1}}
\end{equation}
where $R_{0}$ is the rate at the fluence threshold of ${\cal F}_{0}$, and used Parkes and ASKAP FRB samples to calculate a power law index consistent with a non-evolving Euclidean distribution $(\alpha = -1.5)$.
If we take $R_{0} = 12.7$\,sky$^{-1}$\,d$^{-1}$ and ${\cal F}_{0}=56.6$\,Jy\,ms, which was derived from the ASKAP FRB sample \citep{james19b}, and scale the fluence from 1.3\,GHz to MWA frequencies using an assumed spectral index of -1.5 \citep{macquart19}, then this value of $R_{0}$ corresponds to a fluence of 1054.3\,Jy\,ms at 185\,MHz.
Therefore using Equation~\ref{eq:frb}, we estimate a rate of 
$R=5.9-31.9$\,sky$^{-1}$\,d$^{-1}$ for FRBs with a fluence ranging from  $>1750$\,Jy\,ms down to $>570$\,Jy\,ms (as a function of increasing DM between $200-3000$\,pc\,cm$^{-3}$) at 185\,MHz.
However, \citet{james19a} also determined that there may be a steepening in the source-count distribution to $\alpha = -2.2$ in the fluence range of $R=4.2-49.1$\,sky$^{-1}$\,d$^{-1}$ for FRBs with a fluence ranging from  $>1750$\,Jy\,ms down to $>570$\,Jy\,ms at MWA frequencies.

These source counts do suggest that there could be FRBs at such a high fluence, and if that is the case then the MWA could potentially detect them using image dedispersion.
However, this analysis is an over simplification since such bright events have yet to be detected at $\sim1$\,GHz, and to apply these statistics at 185\,MHz we are assuming that the FRB emission is broadband, extending down to MWA frequencies.
Currently, we have have little information about the behaviour of FRBs at MWA frequencies other than there might be a spectral turn-over of ASKAP-detected FRBs above 200\,MHz \citep{sokolowski18}.
The observation of FRBs down to 400\,MHz by \citet{CHIME2019a} does indicate that some FRBs will be visible at the MWA frequency range.

The triggered MWA observations of SGRBs are also specifically designed to target several SGRB coherent emission models that predict signals within the first 30\,min post-burst \citep{rowlinson19}.
Many of these models depend on the formation of a stable or unstable magentar, which is evidenced by ongoing energy injection in the \swift{} X-ray light curves of many SGRB events \citep{rowlinson13}. In the following, we use the X-ray light curve to model the parameters of the magnetar formed following GRB 180805A, which are then used in conjunction with our fluence and flux density upper-limits to constrain the coherent radio emission from this merger.

\subsection{Central engine activity}\label{sec:mag}
\begin{figure}
    \centering
    \includegraphics[width=\linewidth]{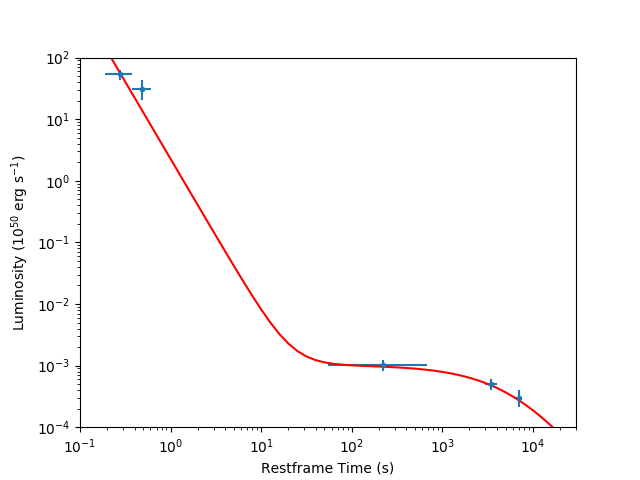}
    \caption{Stable magnetar model fit to the \swift-BAT and \swift-XRT light curve of GRB 180805A following formalism by \citet{rowlinson13}.}
    \label{fig:magfit}
\end{figure}

In order to constrain the central engine activity of GRB 180805A, we obtained the 0.3--10 keV energy band light curve from the {\it Swift} Burst Analyser \citep{evans10}.
This light curve comprises the gamma-ray data obtained by the \swift-BAT and the unabsorbed X-ray data from \swift-XRT.
Using the average short GRB redshift ($z = 0.7$), a k-correction \citep{bloom01}, and assuming the spectrum can be described as a single power law, we create a rest frame 1--10,000 keV luminosity light curve as plotted in Figure~\ref{fig:magfit}.
The light curve of GRB 180805A shows energy injection up to at least $10^{3}$\,s following the merger as evidenced by the late-time plateau phase beginning at $\sim30$\,s post-burst. 

The rest frame light curve was fitted with the magnetar central engine model \citep{zhang01}, following the method outlined in \cite{rowlinson13}.
The fitted model suggests that a stable magnetar might have formed with a magnetic field strength of $B = 2.68f^{+1.62}_{-1.25}\times10^{15}$\,G and a spin-period of $P = 5.16f^{+1.22}_{-1.25}$\,ms, where $f = \left(\frac{\epsilon}{1-\cos\theta}\right)^{0.5}$ encompasses the uncertainties in the efficiency, $\epsilon$, and the beaming angle, $\theta$, of the emission.
These magnetar properties are used to constrain the expected coherent radio emission that may have been produced by GRB 180805A in Section~\ref{sec:coherent}.

\subsection{Coherent emission models}\label{sec:coherent}

If the X-ray light curve of GRB 180805A suggests that a stable magnetar may have been formed following the SGRB event, then we expect that coherent radio emission could have been produced by this source either just prior, during or shortly following the merger.
Using these magnetar parameters, we are able to make predictions about the expected flux density of any associated coherent radio emission following the prescription outlined in \citet[][note that these models still remain highly uncertain]{rowlinson19}.
Given the rest frame X-ray light curve, there are three key models that we are able to test:
\begin{enumerate}[label = {\alph*.}]
    \item \textit{Magnetic field alignment of merging neutron stars} \citep[e.g][]{lyutikov13}. Following \cite{rowlinson19}, the flux density, $F_{\nu}$, is predicted to be
    \begin{equation}
        F_{\nu} \sim 2\times10^{8} (1+z) \frac{B_{15}^2 M_{1.4}^3 }{R_6 \nu_{9, {\rm obs}} D^2} \epsilon_r ~{\rm Jy}    
    \end{equation}
    where $B_{15} = \frac{B}{10^{15}~{\rm G}}$ is the magnetic field of one of the merging neutron stars, $M_{1.4} = \frac{M}{1.4 M_{\odot}}$ is the mass of one of the merging neutron stars, $R_6 = \frac{R}{10^{6} ~{\rm cm}}$ is the radius of one of the merging neutron stars, $\nu_{9,obs}$ is the observing frequency in GHz, $D$ is the distance to the binary system in Gpc and $\epsilon_{r}$ is the fraction of the wind luminosity that is converted into radio emission. Here we assume that the two neutron stars are of standard mass and size ($M_{1.4}=R_6=1$), the magnetic field is $B_{15}=10^{-3}$ and either the typical pulsar value of $\epsilon_r = 10^{-4}$ \citep{taylor93} or a much lower efficiency of $\epsilon_r = 10^{-6}$ \citep{2006ApJ...643..332F}.
    
    \item \textit{Interaction between the relativistic jet and the surrounding medium} \citep{usov00}.
    \citet{usov00} suggest that the jet may be dominated by a Poynting flux wind and that its properties are directly linked to that of the newly formed magnetar that is powering the GRB, and therefore linked to the luminosity  
    of the coherent radio pulse at low frequencies \citep[see prescription and assumptions in section 2.4 of][]{rowlinson19}.
    The fluence of the coherent radio emission at frequency $\nu$ is
    \begin{equation}\label{eq:jet_flu}
        \Phi_{\nu} \simeq \frac{0.1 \epsilon_{B} (\beta - 1)}{\nu_{\rm max}} \left( \frac{\nu_{\rm obs}}{\nu_{\rm max}} \right)^{-\beta} \Phi_{\gamma} ~{\rm erg ~cm}^{-2} {\rm Hz}^{-1} 
    \end{equation}
    for $\nu > \nu_{\rm max}$ (this is the expected regime at MWA frequencies), where $\nu_{\rm max}$ is the peak frequency of the coherent radio emission, $\epsilon_B$ is the fraction of the wind energy contained in the magnetic field (assumed to be $10^{-3}$), $\beta \simeq 1.6$ is the spectral index of the emission, and $\Phi_{\gamma}$ is the observed fluence in gamma-rays in erg\,cm$^{-2}$ \citep[$1.1 \pm 0.3 \times 10^{-7}$\,erg\,cm$^{-2}$ in the $0.3-10$\,keV energy band for GRB 180805A;][]{palmer18}.  
    $\nu_{\rm max}$ is dependent on the magnetic field at the shock front, and therefore the spin period and magnetic field strength of the magnetar remnant \citep{usov00}.
    As the magnetar remnant properties are derived from the X-ray luminosity and light curve in the rest-frame (see Section~\ref{sec:mag}), these values will change with redshift causing the overall radio fluence prediction to also change in brightness over redshift \citep{rowlinson19}. 
    \item \textit{Persistent pulsar-like (dipole spin-down) emission from a magnetar remnant}  \citep[e.g.][]{totani13}. 
    Shortly following the merger, the magnetic field and rotational axes of the magnetar are likely to be aligned, and are therefore along the same axis as the GRB jet, which is our line-of-sight \citep[see discussion by][and references therein]{rowlinson19}.
    The predicted pulsar-like emission for this model is given by
    \begin{equation}\label{eq:pers}
        F_{\nu} \simeq 8\times10^{7} \nu_{\rm obs}^{-1} \epsilon_{r} D^{-2} B_{15}^{2} R_6^6 P_{-3}^{-4} ~{\rm Jy} 
    \end{equation}
    where $P_{-3}$ is the spin period in milliseconds and $B_{15}$ in 10$^{15}$ G is the magnetic field of the newly formed magnetar, which were calculated for GRB 180805A in Section~\ref{sec:mag}. Again, we assume either a typical pulsar value of $\epsilon_r = 10^{-4}$ or a much lower efficiency of $\epsilon_r = 10^{-6}$.
\end{enumerate}

\begin{figure}
    \centering
    \includegraphics[width=\linewidth]{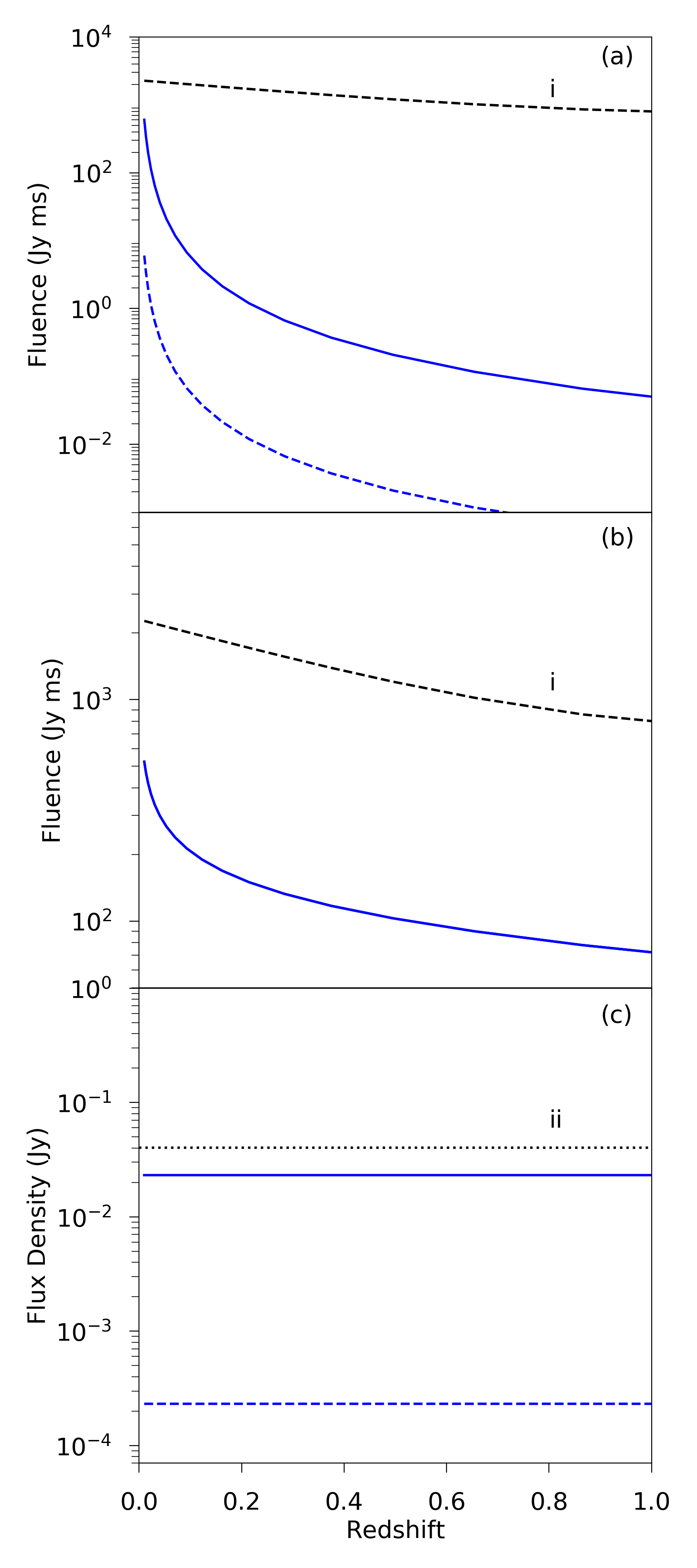}
    \caption{GRB 180805A constraints on the predicted fluence and flux density of three coherent radio emission models as a function of redshift.
    a) Coherent signal emitted prior to the merger from the alignment of magnetic fields for a standard pulsar efficiency of $\epsilon_r = 10^{-4}$ (solid blue curve) or an efficiency of $10^{-6}$ (dashed blue curve). The dashed black line labelled (i) shows the MWA fluence limit ($6\sigma$) as a function of redshift based on the dedispersion analysis of GRB 180805A (see Section~\ref{sec:ided}). 
    b) The relativistic jet interaction with the ISM. The radio fluence predicted using the magnetar parameters derived from the X-ray light curve in Section~\ref{sec:mag} and the gamma-ray fluence of GRB 180805A (solid blue curve).
    Here, we assume the fraction of wind energy in the magnetic field to be $\epsilon_{B} = 10^{-3}$. The black dashed line is the same as for a). 
    c) Persistent pulsar-like emission where the solid and dashed blue lines are the same as for a). The black dotted line labelled (ii) shows the MWA 40.2\,mJy 3$\sigma$ flux density limit from the full 30 min observation of GRB 180805A.}
    \label{fig:coherent}
\end{figure}

As with figure 7 in \cite{rowlinson19}, we can scale these predicted fluence and flux density values to a range of redshifts, which are plotted in Figure~\ref{fig:coherent}.
Each plot panel is labelled a) to c), which corresponds to the three models described above. For panels a) and b), we plot the fluence of the predicted emission as we expect these signals to be prompt (FRB-like).
For model a) the prompt signals are assumed to have a duration of 10\,ms so we multiply the flux density prediction in Equation 5 by this pulse width to obtain the predicted fluence.
The fluence sensitivity limit as a function of DM (and therefore redshift) calculated from the de-dispersed 0.5\,s images at the position of GRB 180805A are also plotted in panels a) and b) to illustrate the constraints placed by this experiment.
The fluence limit attained in the images is simply the limiting flux density multiplied by the integration time of the image in milliseconds. 
Similarly, the $3\sigma$ upper-limit at the position of GRB 180805A calculated from the mean RMS of the the full 30\,min MWA triggered observation is plotted in panel c).

Due to the rapid response of the MWA observations of GRB 180805A, we were on-target to constrain the expected signals from the alignment of the NS magnetic fields prior to the merger \citep[model proposed by][]{hansen01,lyutikov13}, where the emission is pulsar-like, with the peak luminosity being dependent on typical neutron star parameters \citep[see Section 2.3.1 of][]{rowlinson19}. Assuming a typical pulsar efficiency of $\epsilon_{r} = 10^{-4}$ \citep{taylor93}, we would not expect to have detected such a signal even if this event was at a low redshift of z < 0.2 (see solid blue curve in Figure~\ref{fig:coherent}a). However, for a lower efficiency, this signal would certainly go undetected for reasonable SGRB redshifts (e.g. $\epsilon_{r} = 10^{-6}$; dashed curve in Figure~\ref{fig:coherent}a). The cause of the expected non-detection is due to the relatively long integration time (0.5 seconds) compared to the expected signal duration of order a few milliseconds. Thus such a signal would be diluted over the 0.5 second integration and bandwidth of the MWA. In order to detect this signal, we likely need to perform observations at a much higher temporal resolution \citep[this is possible using the MWA Voltage Capture System or VCS;][]{tremblay15}.
High time resolution data processing from the VCS would allow for coherent beamforming at the GRB location (or a range of locations if the GRB localisation covers a larger area).

Similarly, the rapid follow-up allows us to probe for emission from the jet-ISM interaction. 
The timescale of this pulse is expected to be equal to or less than the duration of the prompt gamma-ray emission phase (which is 1.68\,s for GRB 180805A). 
Our fluence limit for GRB 180805A is not constraining for any of the redshift range. 
However, many \swift{} SGRBs have been detected with gamma-ray fluences $>10^{-6}$\,erg\,cm$^{-2}$ \citep[an order of magnitude greater than GRB 180805A, for example see][]{davanzo14} and given that the radio fluence linearly scales with the gamma-ray fluence \citep[see Equation~\ref{eq:jet_flu} and][]{usov00}, a more powerful event could be detectable in our MWA observations, particularly if it is at a low redshift.  
Note that for this prediction, we assume that the fraction of the energy available in the magnetic fields is $\epsilon_{B} = 10^{-3}$ and this factor is currently very poorly constrained for GRBs, with values ranging from $10^{-6}$ -- $10^{-3}$ \citep[e.g.][]{satana2014}, with lower values of $\epsilon_{B}$ further reducing the radio fluence predictions.
This value, and the properties of the newly formed magnetar, are expected to vary between SGRBs and hence further rapid response observations of multiple GRBs may lead to future detections.

Finally, if a magnetar was formed, we would expect pulsar-like dipole radiation to be produced during its lifetime \citep{totani13}.
We would expect such emission to be beamed in a similar way to pulsars, and therefore most likely along the direction of the SGRB jet.
If the jet axis remains pointed towards the Earth, the radio emission would likely be persistent, yet rapidly fading as the magnetar spins down.
Once again, the predicted radio flux density is dependent on the magnetar parameters derived in Section~\ref{sec:mag}.
If the jet-axis became slightly misaligned with the spin axis of the magnetar, it could possibly form a repeating FRB \citep{metzger17}. 
Following the prescription in section 2.5.1. of \citet{rowlinson19}, and for a typical pulsar efficiency of $\epsilon_{r} = 10^{-4}$ (solid blue line in Figure~\ref{fig:coherent}c), this persistent emission is expected to be just below the detection threshold of the deep MWA Phase II 30\,min observation of GRB 180805A (black dotted line in Figure~\ref{fig:coherent}c).
For lower pulsar efficiencies, the 40.2\,mJy threshold for GRB 180805A is extremely unconstraining (dashed blue line in  Figure~\ref{fig:coherent}c).
We note that these predictions are highly dependent upon the magnetar spin period and the magnetic field strength (c.f. Equation~\ref{eq:pers}). The magnetar fitted to the X-ray data of GRB 180805A has a relatively low magnetic field and a long spin period, which makes the predicted coherent radio emission fainter.
The majority of magnetars fitted to the population of SGRBs have higher magnetic fields and faster spin periods \citep{rowlinson13}, hence future events are more likely to produce detectable emission \citep[see the predictions in figure 9c of][]{rowlinson19}.

Since the modelling of the X-ray light curve shown in Section~\ref{sec:mag} indicates that a stable magnetar may have formed following GRB 180805A, this rules out the possibility of a coherent radio pulse being emitted by the collapse of an unstable magnetar into a black hole \citep{falcke14,zhang14}. In addition, as the triggered MWA observations of GRBs are usually 30\,minutes in duration, this coherent emission may be missed in future GRB triggers as the total integration time may be shorter than the magnetar collapse timescale. 

Given that no prompt or coherent radio emission was detected from GRB 180805A over the DM range of $200-3000$\,pc\,cm$^{-3}$ using a light curve transient analysis or via image dedispersion, we can infer from Figures~\ref{fig:coherent}a, \ref{fig:coherent}b and \ref{fig:coherent}c that the predicted coherent radio emission for this GRB, even in the most optimistic scenarios, lie below the MWA detection threshold. Therefore these models are poorly constrained by this GRB, as a detection would only be possible if the efficiency of producing the coherent radio emission is significantly larger than that of normal pulsars.  
It is also worth mentioning that propagation effects that absorb or scatter coherent radio emission may also hinder the detectability of a signal \citep[see e.g.][for a discussion of these propagation effects]{rowlinson19}.
However, we note that for future SGRBs, a higher flux X-ray plateau phase will lead to correspondingly brighter coherent radio emission for all three explored models that could be detectable by our MWA rapid-response observations  
\citep[see analysis by][]{rowlinson19}. 

\subsection{Comparison to other facilities}\label{sec:comp_fac}

The majority of new-generation low-frequency facilities are searching for coherent radio emission associated with GRBs (both short and long) and gravitational wave (GW) events.
In the following, we compare the sensitivities and follow-up response times of these different facilities and observing modes to the MWA triggered observations of GRB 180805A, as well as the other published MWA triggered SGRB \citep[GRB 150424A;][]{kaplan15}, and discuss their strengths and limitations for the study of prompt emission associated with GRBs and GW events.

\begin{figure}
    \centering
    \includegraphics[width=\linewidth]{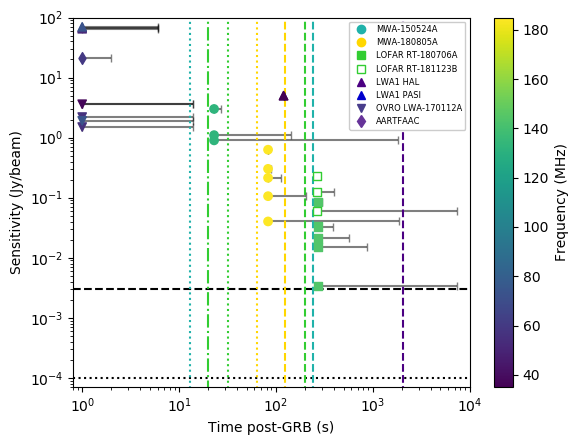}
    \caption{Comparison between the sensitivity and response time for different low-frequency telescopes. 
    The plotted points give the sensitivity for different integration times, which are represented by the black bars, and are colour coded based on the central observing frequency. 
    Different symbols are used to represent the different telescopes. 
    Those plotted at 1\,s post-burst are continuous monitoring programs (the majority all-sky) that obtain simultaneous observations of GRBs.
    The MWA, LOFAR-RT, LWA1 PASI, and OVRO-LWA limits are from the follow-up of real GRBs. 
    The dotted and dot-dashed vertical lines show the arrival time post-burst of the \swift-BAT GRB alerts for the MWA (GRB 150424A: dark green and GRB 180805A: yellow) and LOFAR (GRB 180706A and GRB 181123B: light green dotted and dot-dashed, respectively) triggered events with their colours matching the frequency of the observation. 
    The dashed vertical lines show the expected dispersion-delayed arrival time of a prompt (FRB-like) 
    signal emitted at the time of the GRB event located at a redshift of $z=0.7$ for the four different central observing frequencies for the rapid-response low-frequency telescopes (185, 144, 132.5 and 45.45\,MHz), assuming a typical Galactic DM contribution of 90\,pc\,cm$^{-3}$. 
    The black horizontal dashed and dotted lines show the predicted best case 2\,hr $3\sigma$ thermal noise limit for the MWA at 185\,MHz \citep{wayth15} and the 2\,hr integration sensitivity ($3\sigma$) for LOFAR at 150\,MHz (using the LOFAR Image noise calculator, assuming a frequency coverage between 120--168\,MHz and that all core, remote and international stations participated in the observation).}
    \label{fig:comp}
\end{figure}

In Figure~\ref{fig:comp}, we show the $3\sigma$ upper-limits of low-frequency telescopes that perform continuous monitoring programs or rapid-response follow-up observations of GRBs. The plotted points give the $3\sigma$ sensitivity for each instrument as a function of the telescope rapid-response-time post-burst (we assume that the \swift{}-BAT detection time of the GRB is its start time at post-burst=0) on different timescales (represented by the black horizontal error bars). The data points are colour-coded based on the central observing frequency (see the colour-bar in Figure~\ref{fig:comp}). 
We plot the sensitivities for GRB 180805A at 185\,MHz on the 0.5\,s, 5\,s, 30\,s, and 2\,m timescales, which correspond to 3 times the average RMS values over the first 2\,min snapshot (ObsID 1217495184). The 30\,min sensitivity was measured from the mean image of all 15 snapshot observations as output by \robbie{}.
The sensitivities plotted for GRB 150424A are at the central observing frequency of 132.5\,MHz, correspond to 4\,s, 2\,m and 30\,m timescales \citep{kaplan15}.
The dotted vertical lines indicate the arrival time of the \swift{}-BAT  VOEvents associated with GRB 180805A (yellow) and GRB 150424A (dark green; the colours correspond to the observing frequency of each GRB).
The MWA sensitivity limits for GRB 180805A are more constraining as they were taken over a bandwidth of 30.72\,MHz as compared to a 2.56\,MHz bandwidth for GRB 150424A, which was observed in a ``picket-fence'' (sub-band) mode.

Other low frequency radio telescopes running rapid-response systems include the recently commissioned LOFAR Responsive Telescope (RT), which uses a similar VOEvent triggering protocol to the MWA through the utilisation of the 4 Pi Sky tools \citep{rowlinson19grb,rowlinson20pp}. On receiving a transient alert, the LOFAR RT triggers the LOFAR High Band Array (HBA; 120--168\,MHz), which has a temporal resolution of 1\,s, and is capable of being on-target within 4-5\,m \citep{rowlinson19grb,rowlinson20pp}. 
In Figure~\ref{fig:comp}, we plot the $3\sigma$ upper-limits from the LOFAR-RT triggered observation of the long GRB 180706A, which began 4.5\,m post-burst, searching for coherent emission on 30\,s, 2\,m, 5\,m, 10\,m and 2\,h timescales \citep{rowlinson19grb}. 
We also include the LOFAR RT results for the short GRB 181123B, which began 4.4\,m post-burst, plotting the $3\sigma$ upper-limits for the 8\,s, 138\,s and 2\,hr timescales.
For both GRBs, we plot the arrival time of the \swift{}-BAT VOEvents  in light green (based on the central observing frequency of 144\,MHz) as a dotted (GRB 180706A) and dot-dashed (GRB 181123B) vertical line. 
The LWA1 also has a rapid-response mode called the Heuristic Automation for LWA1 (HAL), which can form up to four steerable beams that can be on-target as fast as two-minutes post-trigger, which can be tuned to 25.85 and 45.45\,MHz \citep{yancey15}. 
For LWA1 HAL, we assume an approximate sensitivity of 5\,Jy \citep[$3\sigma$;][]{taylor12,ellingson13} for a 1\,s integration at 45.45\,MHz. 

In order to compare the abilities of these three low-frequency triggering instruments, we also depict the expected dispersion-delayed arrival times of a  prompt/coherent radio signal at four different observing frequencies (chosen to be relevant for triggered observations with the MWA, LOFAR RT and LWA1 HAL) as vertical dashed lines in Figure~\ref{fig:comp}, assuming they were emitted at the time of the GRB for an average SGRB redshift of $z=0.7$.
The dispersion-delayed arrival times also assume a typical Galactic DM of 90 pc\,cm$^{-3}$ \citep[185\,MHz - yellow, 144\,MHz - light green, 132.5\,MHz - dark green, and 45.45\,MHz - indigo, see equation 2 in][]{james19c}.
This clearly demonstrates that both the MWA and LWA1 HAL are capable of being on-target before the arrival of any prompt radio signals emitted at the time of the GRB at this average redshift for the specified observing bands.
Given the LOFAR RT trigger delay is between $\sim4$ to $5$\,m, it will not be on-target in time to detect such a signal emitted during or just prior to the GRB at 144\,MHz. 
The advantage of LOFAR RT is its sensitivity, which makes it more relevant for probing persistent, coherent radio emission that could be emitted for up to several hours following the GRB event by its magnetar remnant. 
However, note that in the case of short GRB 181123B, LOFAR showed similar sensitivities on short timescales (between 8 and 140\,s) to the MWA triggered observations of GRB 180505A \citep{rowlinson20pp}. 
LOFAR RT triggers also observe for up to 2\,h (as opposed to the 30\,m observation obtained by the MWA) so is more likely to detect any prompt emission emitted by the collapse of said magnetar (if unstable) into a black hole \citep[see discussions in][]{rowlinson19}.

There are also several low frequency facilities with all-sky and/or continuous monitoring capabilities that rely on serendipitous observations of cataclysmic and merger events. These include the the LOFAR all-sky monitor called the Amsterdam ASTRON Radio Transient Facility and Analysis Centre \citep[AARTFAAC;][]{prasad14,prasad16}, which operates when the LOFAR Low Band Array (LBA; 10-90\,MHz) is observing. The Owens Valley Radio Observatory Long Wavelength Array \citep[OVRO-LWA; e.g.][]{eastwood18} is another example of an all-sky monitor. Its transient buffer can buffer up to 24\,h of data with a temporal and spectral resolution of 13\,s and 24\,kHz, respectively, between 24-84\,MHz. Using this transient buffer, both \citet{anderson18lwa} and \citet{callister19} have placed similar limits on coherent emission from the short GRB 170112A and the binary black hole merger GW170104, respectively. The first station of the Long Wavelength Array \citep[LWA1, 10-88\,MHz;][]{taylor12,ellingson13} also operates two all-sky transient observing modes, one of which is equipped with the Prototype All Sky Imager \citep[PASI;][]{obenberger15} backend, which images in near-real-time at the native temporal resolution of 5\,s. LWA1-PASI has already serendipitously observed GRBs, placing limits at 37.9, 52.0 and 74\,MHz \citep{obenberger14}. The sensitivities and corresponding temporal resolution of these monitoring, all-sky instruments are included in Figure~\ref{fig:comp} at time 1\,s post-burst. While by definition these three programs will be on-target in-time to receive associated prompt, coherent radio emission associated with a GRB, they are unlikely to reach the sensitivities required to detect them \citep{rowlinson19}. The biggest limitation is their temporal resolution, diluting any prompt signals, which are predicted to last for only milliseconds in most cases. Note there are already plans for the OVRO-LWA buffer to be upgraded to record at a much finer temporal resolution, which will make them more sensitive to prompt radio transients \citep{callister19}. 

Overall, the MWA is the most competitive low frequency radio telescope for detecting prompt signals emitted by a GRB just prior-to or during the merger. This is due to its rapid-response triggering mode, combined with its competitive instantaneous $(u,v)$ coverage and sensitivity ($0.5$\,s) when compared to all other facilities that are capable of being on-target either during or rapidly following the GRB alert.

\begin{figure}
    \centering
    \includegraphics[width=\linewidth]{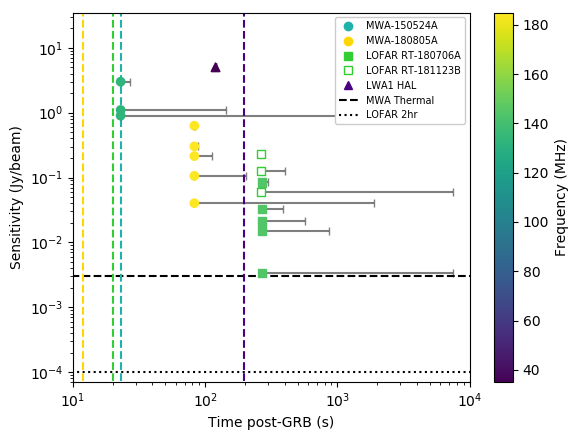}
    \caption{As for Figure~\ref{fig:comp} but the dashed vertical lines show the expected arrival time of prompt emission associated with a BNS merger at the distance of GW170817 \citep[40\,Mpc;][]{abbott17b} for four different central observing frequencies (185, 144, 132.5 and 45.45\,MHz), assuming a typical Galactic DM contribution of 90\,pc\,cm$^{-3}$.}
    \label{fig:compgw}
\end{figure}

\subsubsection{Implications for rapid GW follow-up}\label{sec:imp_gw}

One source class for which the standard MWA triggered observations are limited is for nearby merger events.
The aLIGO/Virgo O3 observing run had a BNS horizon limit of $\sim170$\,Mpc, which is far closer than the redshift range of \swift-detected SGRBs.
While associated FRB-like emission will be brighter due to the event's proximity, an emitted prompt radio signal will be far less dispersion delayed. 
Taking into account the delays associated with transmitting the alert (best case for aLIGO/Virgo is $\sim10-20$\,s) and the rapid-response repointing time of the MWA ($4-16$\,s), the telescope is unlikely to be on-target fast enough for observing frequencies $\gtrsim136$\,MHz \citep[assuming a typical Galactic DM of 90\,pc\,cm$^{-3}$;][]{james19c}. This is demonstrated in Figure~\ref{fig:compgw}, which shows the response times and sensitivities of the low frequency rapid-response telescopes (MWA, LOFAR RT and LWA1 HAL, similar to Figure~\ref{fig:comp}) but with the expected arrival time of a coherent, prompt signal emitted at the moment of a BNS merger event located at the distance of GW170817 \citep[40\,Mpc;][]{abbott17b}. Again, we assume a typical Galactic DM of 90\,pc\,cm$^{-3}$ when calculating the signal dispersion delay, however, this value can vary greatly with the Galactic latitude of the event \citep[see figure 1 of][]{hancock19b} and may therefore have a much shorter dispersion delay. 
In Figure~\ref{fig:compgw}, it can be seen that rapid-response observations taken at the lower end of the MWA observing band ($\lesssim130$\,MHz) and observations within the LWA1 HAL observing band may be on-target in-time to detect associated prompt signals. However, with the exception of the recent LOFAR 110-188 MHz detections of the repeating FRB 180916B \citep{pleunis21},
prompt, coherent radio signals (aka FRBs) have not been detected below 400\,MHz \citep[e.g. see][]{tingay15,rowlinson16,sokolowski18,houben_constraints_2019,terVeen_FRATS_2019}, which motivates searches at frequencies $>130$\,MHz. 

Despite the shorter dispersion delays, the MWA still provides the best chance of detecting prompt, coherent emission associated with aLIGO/Virgo-detected BNS mergers due to its superior instantaneous sensitivity and rapid repointing capabilities.
\citet{james19c} therefore devised a strategy for using the MWA to trigger on `negative latency' aLIGO/Virgo alerts of BNS mergers, which involves a trigger being generated via the detection of gravitational waves associated with the inspiral of the compact objects rather than waiting for the signal produced by the merger.
This necessarily shrinks our BNS merger aLIGO/Virgo horizon with negative latency as inspiral GWs are fainter compared to the merger.
However, the precious seconds gained can be enough time for the MWA to begin observing the event before the arrival of any prompt, coherent radio emission. For further details on the proposed observing mode, see \citet{james19c}.

Given the field of view of the MWA is $\sim 1000\, \mathrm{deg}^2$ for the most commonly used observing frequencies ($120-200$\,MHz) there is also a small ($\sim 2-3\%$) chance that any GW/SGRB that occurs during an MWA observation will be captured by the MWA without any need for triggering or re-pointing of the instrument.

\subsection{Prospects for improvement}

Since the automated triggered observations were performed for this GRB, we have continued to improve the response time of the MWA triggering system.
We have optimised the front end processing of the VOEvents to reduce the latency between receiving an event and requesting an observation from 2.8s (event 3 in Table\,\ref{tab:trigger}) to just 4\,ms.
For this work, all the GRB observations were queued at once, and were therefore not queued until the Sun avoidance calculation was performed \citep[see Section 3.1.2 of][]{hancock19b}. 
By computing the Sun avoidance for only the first observation, queuing that observation immediately, and then computing and queuing the remainder of the 14 observations, it is possible to further reduce the time between the initial trigger and the time of the first observation by an additional 1.7\,s.

An overall improvement to this experiment is to use the Voltage Capture System \citep[VCS;][]{tremblay15} when triggering rapid-response observations with the MWA due to its high temporal and spectral resolution of $100\,\mu$s and $10$\,kHz, respectively. 
Such a temporal resolution will improve our sensitivity to prompt (likely millisecond timescale) coherent emission, rather than the signals being smeared by the standard correlator temporal resolution of 0.5\,s. 
The higher frequency resolution will also make us more sensitive to determining the lower DM of nearby events, which is particularly relevant for GW triggers as aLIGO/Virgo sensitivities only allow for searches of neutron star mergers out to $<200$\,Mpc \citep{buikema20}.
In fact, \citet{rowlinson19} made predictions about prompt coherent emission from SGRBs over a wide range of redshifts, and compared the capabilities of many of the low frequency radio telescopes that are actively targeting these events (e.g. see Figure~\ref{fig:comp}) to detecting such signals.
They demonstrated that at early times (seconds to minutes post-burst/merger) the MWA VCS is the most competitive instrument for making such detections.  
VCS triggering on transient alerts in the form of VOEvents have also recently been commissioned for the MWA \citep[see][]{hancock19b}. 

Another suggested improvement will likely come with the MWA Phase III, when we anticipate that all 256 MWA tiles will be capable of observing simultaneously (currently only 128 can be run at any one time).
Observing with all 256 tiles will improve the instantaneous sensitivity of MWA and therefore make it more sensitive to short timescale emission.
Expanding the Sun avoidance component of the rapid-response code to allow for other known bright sources (e.g. the A-team sources) to be placed in a primary beam null would also improve the sensitivity of the triggered observations.
This would have been advantageous for our rapid-response observations of GRB 180805A as Cen A was moving into the primary beam sidelobe within the first 10\,min of the observation. 
Having a more complete sky model that includes good solutions to the brighter sources would also help with providing a better calibration, and therefore better sensitivity. Such a sky model is being developed as part of the GLEAM-X processing pipeline.\footnote{https://github.com/nhurleywalker/GLEAM-X-pipeline}

\section{Summary and Conclusions}

We present the first triggered observation of a SGRB using the upgraded MWA rapid-response mode. The MWA was observing GRB 180805A 83.7\,s post-burst and was therefore on-target to probe several models (see Section~\ref{sec:coherent}) that predict prompt, coherent radio signals to be emitted either just before or for several minutes following a merger event. We searched for such signals using three different methods, including: 
\begin{enumerate}
    \item A transient search at the \swift-XRT position of GRB 180805A over 0.5\,s, 5\,s, 30\,s and 2\,min timescales (resulting in $3\sigma$ flux density limit ranges of $570-1830$, $270-630$, $200-420$, and $100-200$\,mJy\,beam$^{-1}$, respectively), which cover the dispersion delay expected across the MWA observing band for a range of known SGRB redshifts.
    \item A search for persistent coherent emission at the \swift-XRT position of GRB 180805A from a (quasi-) stable magnetar remnant over the full 30\,min MWA observation ($3\sigma $ flux density upper-limit of 40.2\,mJy\,beam$^{-1}$).
    \item A search for dispersed radio transients at the \swift-XRT position of GRB 180805A in the image domain using 0.5\,s coarse channel (24 times 1.28\,MHz) images, which tested for pulse start times and DMs in steps of 0.1\,s and 1\,pc\,cm$^{-3}$, respectively, over the DM range of 200-3000\,pc\,cm$^{-3}$. 
    This resulted in a $6\sigma$ fluence upper-limit range from 570\,Jy\,ms at DM$=3000$\,pc\,cm$^{-3}$ to 1750\,Jy\,ms at DM$=200$\,pc\,cm$^{-3}$. 
    Given how rapidly the MWA was on target and the dispersion delay of such signals at MWA observing frequencies, we were sensitive to any prompt emission emitted at the time of the GRB 180805A (i.e. when the first gamma-rays were emitted) for DM trial values $\gtrsim800$\,pc\,cm$^{-3}$, and therefore up to 350\,s before the merger for DMs between 800-3000\,pc\,cm$^{-3}$. 
\end{enumerate}

No transients were detected at the position of GRB 180805A using any of the above methods, yet we place some of the most stringent constraints on prompt, dispersed, coherent radio emission on an SGRB to-date on short timescales and using image dedispersion techniques. 
Results using these transient analysis techniques to search for coherent radio signals from a larger sample of SGRBs that were triggered between 2017 and 2020 using the MWA rapid-response system will appear in the companion paper Tian et al., in prep.

Despite the deep limits obtained for this GRB, we were unable to place meaningful constraints on three models that predict coherent radio emission from SGRBs.
Specifically, we cannot constrain the models predicting emission from the magnetic field alignment of the merging neutron stars \citep{lyutikov13}, 
from the interaction between the gamma-ray jet and the ISM \citep{usov00}
or from persistent pulsar-like emission from the magnetar remnant \citep{totani13}. 
However, future events hold significant promise as with a more powerful GRB with a gamma-ray fluence an order of magnitude higher than GRB 180805A and/or 
a brighter X-ray counterpart we would expect a brighter prompt radio signal from the jet-ISM interaction and potentially a detectable pulsar-like radio counterpart.
In addition, these models could be more tightly constrained for a GRB with a known redshift \citep{usov00}.
Finally, if such an event were then at a low redshift ($z\lesssim0.5$) then according to these models our MWA observations may be sensitive enough to detect this coherent emission using the image dedispersion technique presented in this paper.

Due to its extremely rapid transient response time, the MWA is therefore currently the most competitive low frequency instrument for probing prompt and coherent radio emission models that predict the production of such signals just prior, during or shortly following a merger event (see Section~\ref{sec:comp_fac}).
While the MWA has the sensitivity to constrain some models, particularly at low redshifts, better instantaneous sensitivity on shorter timescales, such as will be possible with SKA-Low, are required to truly probe many of these coherent emission models.
This illustrates the importance of ensuring that the SKA-Low has a rapid-response mode of operation that is capable of being triggered and on-target within seconds of receiving a transient alert. 

\section*{Acknowledgements}
{\bf Facilities:}
This scientific work makes use of the Murchison Radio-astronomy Observatory, operated by CSIRO. We acknowledge the Wajarri Yamatji people as the traditional owners of the Observatory site. Support for the operation of the MWA is provided by the Australian Government (NCRIS), under a contract to Curtin University administered by Astronomy Australia Limited.

This work made use of data supplied by the UK {\it Swift} Science Data Centre at the University of Leicester and the {\it Swift} satellite. {\it Swift}, launched in November 2004, is a NASA mission in partnership with the Italian Space Agency and the UK Space Agency. {\it Swift} is managed by NASA Goddard. Penn State University controls science and flight operations from the Mission Operations Center in University Park, Pennsylvania. Los Alamos National Laboratory provides gamma-ray imaging analysis. 

This work was supported by resources provided by the Pawsey Supercomputing Centre with funding from the Australian Government and the Government of Western Australia.

{\bf Funding:}
GEA is the recipient of an Australian Research Council Discovery Early Career Researcher Award (project number DE180100346), and JCAM-J, NH-W, and TM are the recipients of the Australian Research Council Future Fellowship (project numbers FT140101082, FT190100231, and FT150100099) funded by the Australian Government. 

{\bf People:}
We thank the referee for their careful reading of the manuscript and recommendations. 
The authors wish to thank Marco De La Pierre, Paolo Di Tommaso, and Evan Floden, for the well timed Nextflow workshop that enabled the authors to port \robbie{} to Nextflow.

{\bf Software:}
The following software and packages were used to support this work: {\sc mwa\_trigger} \citep{hancock19b}\footnote{\url{https://github.com/MWATelescope/mwa\_trigger/}}, 
GLEAM-X pipeline\footnote{\url{https://github.com/nhurleywalker/GLEAM-X-pipeline}}, 
MWA-fast-image-transients\footnote{\url{https://github.com/PaulHancock/MWA-fast-image-transients}},
{\sc flux\_warp} \citep{duchesne20}\footnote{\url{https://gitlab.com/Sunmish/flux\_warp/}},
{\sc comet} \citep{swinbank14}, 
{\sc voevent-parse} \citep{staley16pp}, 
\robbie{} \citep{hancock19}\footnote{\url{https://github.com/PaulHancock/Robbie}}, 
AegeanTools \citep{hancock18}\footnote{\url{https://github.com/PaulHancock/Aegean}}, 
{\sc fits\_warp} \citep{hurley-walker18}\footnote{\url{https://github.com/nhurleywalker/fits\_warp}}, 
TOPCAT/stilts \citep{taylor05}, 
{\sc Astropy} \citep{TheAstropyCollaboration2013,TheAstropyCollaboration2018}, 
{\sc numpy} \citep{vanderWalt_numpy_2011}, 
{\sc scipy} \citep{Jones_scipy_2001}, 
{\sc matplotlib} \citep{hunter07},   docker\footnote{\url{https://www.docker.com/}}, 
singularity \citep{kurtzer_singularity_2017}.
This research has made use of NASA's Astrophysics Data System. 
This research has made use of SAOImage DS9, developed by Smithsonian Astrophysical Observatory.
This research has made use of the VizieR catalogue access tool \citep{ochsenbein00} and the SIMBAD database \citep{wenger00}, operated at CDS, Strasbourg, France.

\bibliographystyle{pasa-mnras}
\bibliography{papers}

\end{document}